%% file: ms.tex
\documentclass[a4paper,11pt]{article}
\pdfoutput=1 
\usepackage{jcappub}
\usepackage{graphicx}
\pdfoutput=1 
\usepackage{color}
\usepackage[dvipsnames]{xcolor}
\usepackage[utf8]{inputenc}
\usepackage{array}
\usepackage{multirow}

\newcommand{\rr}{\mathrm}

\title{Constraining Dark Matter - Dark Radiation interactions with CMB, BAO, and Lyman-$\alpha$}

\author[1,2]{Maria Archidiacono}
\author[1]{, Deanna C. Hooper}
\author[3]{, Riccardo Murgia}
\author[4]{, Sebastian Bohr}
\author[1]{, Julien Lesgourgues}
\author[3,5]{, Matteo Viel}

\affiliation[1]{Institute for Theoretical Particle Physics and Cosmology (TTK), \\ RWTH Aachen University, D-52056 Aachen, Germany.}
\affiliation[2]{INFN, Sezione di Bologna, viale Berti Pichat 6/2, 40127 Bologna, Italy and\\
Dipartimento di Fisica e Astronomia, Alma Mater Studiorum Universit\`a di Bologna, Via Gobetti, 93/2, I-40129 Bologna, Italy}
\affiliation[3]{SISSA, Via Bonomea 265, 34136 Trieste, Italy \\
INFN, Sez. di Trieste, Via Valerio 2, 34127 Trieste, Italy \\
IFPU, Institute for Fundamental Physics of the Universe, via Beirut 2, 34151 Trieste, Italy}
\affiliation[4]{Center for Astrophysics and Cosmology, Science Institute, \\University of Iceland, Dunhagi 5, 107 Reykjavik, Iceland}
\affiliation[5]{INAF,  Osservatorio Astronomico di Trieste,
via Tiepolo 11, I-34131 Trieste, Italy}

\emailAdd{maria.archidiacono@bo.infn.it}
\emailAdd{hooper@physik.rwth-aachen.de}
\emailAdd{riccardo.murgia@sissa.it}

\abstract{Several interesting Dark Matter (DM) models invoke a dark sector leading to two types of relic particles, possibly interacting with each other: non-relativistic DM, and relativistic Dark Radiation (DR). These models have interesting consequences for cosmological observables, and could in principle solve problems like the small-scale cold DM crisis, Hubble tension, and/or low $\sigma_8$ value. Their cosmological behaviour is captured by the ETHOS parametrisation, which includes a DR-DM scattering rate scaling like a power-law of the temperature, $T^n$. Scenarios with $n=0$, $2$, or $4$ can easily be realised in concrete dark sector set-ups. Here we update constraints on these three scenarios using recent CMB, BAO, and high-resolution Lyman-$\alpha$ data. We introduce
a new Lyman-$\alpha$ likelihood that is applicable to a wide range of cosmological models with a suppression of the matter power spectrum on small scales. For $n=2$ and $4$, we find that Lyman-$\alpha$ data strengthen the CMB+BAO bounds on the DM-DR interaction rate by many orders of magnitude. However, models offering a possible solution to the missing satellite problem are still compatible with our new bounds. For $n=0$, high-resolution Lyman-$\alpha$ data bring no stronger constraints on the interaction rate than CMB+BAO data, except for extremely small values of the DR density. Using CMB+BAO data and a theory-motivated prior on the minimal density of DR, we find that the $n=0$ model can reduce the Hubble tension from $4.1\sigma$ to $2.7\sigma$, while simultaneously accommodating smaller values of the $\sigma_8$ and $S_8$ parameters hinted by cosmic shear data.}

\begin{document}

\hfill{\small TTK-19-23}

\maketitle

\input{Introduction}
\input{Theory}
\input{Methods}
\input{Results}
\input{Conclusions}

\section*{Acknowledgement}
The authors thank N. Sch\"{o}neberg and T. Ronconi for useful discussions. Simulations were performed with computing resources granted by JARA-HPC from RWTH Aachen University under project jara0184, and on the Ulysses SISSA/ICTP supercomputer. MV and RM are supported by the INFN INDARK PD51 grant. DH and JL are supported by the DFG grant LE 3742/3-1. SB is supported by a Grant of Excellence from the Icelandic Research Fund (grant number 173929-051).

\bibliography{biblio}{}
\bibliographystyle{JHEP}

\end{document}

%% file: Introduction.tex
\section{Introduction}
\label{sec:intro}
The standard cosmological model, in which Dark Matter (DM) is cold and collisionless, boasts remarkable success across many different scales. Its ability to simultaneously explain the early-time Cosmic Microwave Background (CMB) anisotropies~\cite{Aghanim:2018eyx} and structure formation at large scales has solidified Cold Dark Matter (CDM) as a cornerstone of modern cosmology.

Despite this, some possible problems on small scales remain unsolved. These come about when comparing N-body simulations of structure formation with observations: firstly, we do not observe as many dwarf satellites as shown in the simulations (missing satellite problem~\cite{Klypin:1999uc}), while the most massive predicted subhalos, which have so much enclosed mass they should have ignited, also remain unseen (too-big-to-fail problem~\cite{BoylanKolchin:2011de}). Furthermore, we observe the density profile of halos to be more core-like than the cuspy profiles preferred by the simulations (core-vs-cusp problem~\cite{Donato:2009ab}), as well showing more diversity in the inner density profile than expected (diversity problem~\cite{Oman2015,Tulin2018}). The inclusion of baryonic feedback is crucial for providing a realistic picture of the aforementioned problems and it shows that baryons can indeed partially solve the CDM crisis~\cite{Garrison-Kimmel:2017zes,Sawala:2015cdf,Pawlowski:2015qta}. However, in the absence of a compelling solution within the $\Lambda$CDM model, alternative scenarios of self-interacting dark matter~\cite{Bellazzini:2013foa,Tulin:2013teo} emerged as a possible way to explain the small scale observations~\cite{Kahlhoefer2019,Zavala:2019sjk,Ren:2018jpt,Robertson:2017hdw,Valli:2017ktb,Tulin:2017ara,Brinckmann:2017uve,Kamada:2016euw,Zavala:2012us,Aarssen:2012fx,Loeb:2010gj}, although the non-trivial interplay between self-interacting dark matter and baryons has to be considered~\cite{Robertson:2018anx}.

In addition to the small scale crisis, the standard cosmological paradigm has also been challenged by possible disagreements in different datasets. The most notable of these is the $H_0$ tension, where the value of the Hubble Constant inferred from CMB and BAO data~\cite{Aghanim:2018eyx}, is significantly lower than the value measured with supernovae~\cite{Riess:2019cxk}. Similarly, measurements of $\sigma_8$ - the amplitude of the matter power spectrum on the scale of $8\,\rr{Mpc}/h$ - also yield a slight tension across different observations; the latest CMB+BAO inferred value~\cite{Aghanim:2018eyx} is slightly higher than the value obtained from weak lensing experiments~\cite{Hildebrandt:2018yau,Abbott:2017wau,Joudaki:2019pmv} (see however~\cite{Fluri:2019qtp}). While many models have been proposed to alleviate the $H_0$~\cite{Agrawal:2019lmo,Poulin:2018cxd,Archidiacono:2016kkh,DiValentino:2016hlg,Efstathiou:2013via} and $\sigma_8$~\cite{SpurioMancini:2019rxy, Murgia:2016ccp} tensions, these are often degenerate with other cosmological effects, making it very hard to solve both simultaneously. 

Moreover, despite the experimental efforts, DM in the form of Weakly Interacting Massive Particles (WIMPs) has so far eluded detection in direct and indirect searches, as well as at colliders~\cite{Bertone:2018xtm}. Together with the potential problems on small scales and the apparent mismatch in different cosmological datasets, this has reinvigorated interest in models beyond the standard CDM paradigm.
 
A class of models that has gained a lot of interest in recent years are those where DM couples to an additional relativistic dark sector, known as Dark Radiation (DR)\cite{Bringmann:2016ilk,Buckley:2014hja,Boddy:2014yra,Buckley:2009in}. Some of these models have been proposed to solve the missing satellite problem~\cite{Archidiacono:2017slj, Vogelsberger:2015gpr, Schewtschenko:2015rno} or to delay reionization \cite{Das18} by means of a cut-off in the matter power spectrum. Other classes of DM-DR models, with a smooth damping of the matter power spectrum, can alleviate the cosmological tensions on $H_0$ and/or $\sigma_8$, as proposed in Refs.~\cite{Buen-Abad:2015ova, Lesgourgues:2015wza, Buen-Abad:2017gxg}.

Given the suppression these interacting DM-DR models can have on small scale structure growth, the matter power spectrum is an essential tool to study these dark sector interactions~\cite{Gluscevic:2019yal}. The Lyman-$\alpha$ forest flux power spectrum, which comes from absorption lines in the spectra of distant quasars due to the gas clouds of neutral hydrogen in the Inter-Galactic Medium (IGM), has been shown to provide very good measurements of the matter power spectrum on the scales significant for these interactions~\cite{Ikeuchi1986, 10.1093/mnras/218.1.25P}, and as such can be used to constrain the properties of DM-DR interactions.

However, to obtain a flux power spectrum for a given cosmological model, detailed knowledge of the intervening hydrogen clouds leading to the Lyman-$\alpha$ forest is needed. This usually requires running computationally expensive hydrodynamical N-body simulations~\cite{bolton17} for every set of underlying cosmological parameters, making MCMC analyses prohibitive. Nonetheless, in Refs.~\cite{Murgia:2017lwo, Murgia:2017cvj, Murgia:2018now} a novel approach was proposed to avoid the need of (many) new simulations.
In this framework the suppression of the matter power spectrum is parametrised using three parameters that are able to capture the full shape of the cut-off. High resolution simulations are performed on a grid of nodes given by several combinations of these parameters, allowing one to interpolate at a later stage on the pre-existing grid to derive constraints on a specific model.

In order to study DM-DR interactions we have developed a new likelihood for the parameter inference code {\sc MontePython}~\cite{Audren:2012vy,Brinckmann:2018cvx} making use of the method proposed in Refs.~\cite{Murgia:2017lwo, Murgia:2017cvj, Murgia:2018now}. This has been used together with our implementation in the Boltzmann code {\sc class}~\cite{Blas:2011rf} of the generic ETHOS parametrisation~\cite{Cyr-Racine:2015ihg} for DM-DR interactions. Here we will present our bounds on DM-DR interactions obtained when using  Lyman-$\alpha$ data from HIRES/MIKE in combination with CMB data from Planck and Baryonic Acoustic Oscillation data. We note that there have been several studies in the past that used the Lyman-$\alpha$ forest data to constrain DM-DR and DM-baryons interactions. In particular, Refs.~\cite{Xu:2018efh,Dvorkin:2013cea, Krall:2017xcw} used measurements of the linear matter power spectrum amplitude and slope obtained from SDSS-II low resolution low signal-to-noise quasar spectra. However, these measurements were obtained assuming a vanilla $\Lambda$CDM cosmology or small departures from it~\cite{McDonald:2004eu,McDonald:2004xn}. More recently, the authors of Ref.~\cite{Garny:2018byk} compared interacting dark sector models to the 1D flux power spectrum derived from SDSS-III data. They proposed a new modelling of the flux power spectrum in which the non-linear evolution is calculated analytically using viscous two-loop perturbation theory \cite{Blas:2015tla}, while uncertainties on the flux power spectrum modelling are accounted for by a marginalizing over several nuisance parameters. None of these investigations rely on a forward modelling of the flux power spectrum based on high resolution hydrodynamical simulations, and this is the approach proposed in this work.

This paper is organised as follows: in section \ref{sec:theory} we present the interacting DM-DR model, and describe its impact on structure formation and cosmological observables. Section \ref{sec:methods} describes the formalism and subsequent implementation of our Lyman-$\alpha$ likelihood, as well as our modifications to the Boltzmann code {\sc class}, while in section \ref{sec:results} we present our main results. Our conclusions and outlook are summarised in section \ref{sec:conc}. 

%% file: Theory.tex
\section{Theoretical framework: Physical effects of Dark Matter - Dark Radiation interactions}
\label{sec:theory}

In order to solve the missing satellite problem, besides self-interactions, DM has to scatter off a relativistic particle. The Standard Model particles (neutrinos and photons) cannot play this role both because of model building issues~\cite{Bringmann:2016ilk} and because of cosmological consequences (e.g. bounds on free-streaming neutrinos~\cite{Archidiacono:2013dua, Brust:2017nmv}). Therefore, we need to invoke the existence of an extra {\it Dark} Radiation (DR) component, which requires an extension of the Standard Model of particles. Here, we will not focus on one specific particle model, because the aim of this paper is to devise a general phenomenological approach that can be applied to several models.

In order to keep the discussion as general as possible, we assume the ETHOS parametrisation~\cite{Cyr-Racine:2015ihg} of DM-DR interactions.
We first implemented the ETHOS framework in {\sc class} in Ref.~\cite{Archidiacono:2017slj} (see section \ref{sec:class} for more details on our implementation), where we studied the effects of DM-DR interactions through a massive mediator on cosmological observables at large scales; in our previous work, we stressed that, given the ETHOS parametrisation, the impact of the specific particle physics model (e.g. the DM mass, the presence of DR self-interactions, the vector or scalar nature of the mediator) have no (or negligible) impact. The only relevant physical quantities are:
\begin{itemize}
\item the temperature dependence of the comoving interaction rate $\Gamma_\rr{DR-DM} \propto T^n$, where $\Gamma_\rr{DR-DM}$ can be seen as the DR drag opacity, i.e. the scattering rate of DR off DM,
\item the strength of the interaction $a_\rr{dark}$ ($\Gamma_\rr{DR-DM} \propto a_\rr{dark}T^n$),
\item the amount of DR parametrised through the temperature ratio $\xi=T_\rr{DR}/T_\gamma$, where $T_\gamma$ is the temperature of CMB photons,
\item the nature of DR (i.e. free-streaming or not).
\end{itemize}

The effects of DM-DR interactions on the CMB were already discussed in detail in Refs.~\cite{Cyr-Racine:2013fsa} and \cite{Archidiacono:2017slj}. Here we summarise the most important differences between such models  and a $\Lambda$CDM model with an equivalent number of extra neutrino-like particles $N_\rr{eff}$ (i.e. with the same background density of radiation):
\begin{itemize}
\item Non-free-streaming DR:
due to its self-interactions and/or its coupling with DM, DR does not lead to additional anisotropic stress, and thus, does not induce the damping and phase-shift of the CMB acoustic peaks that is typically expected in presence of additional relativistic degrees of freedom;
\item Non-growing DM fluctuations: the momentum exchange between DM and DR particles reduces the growth rate of DM perturbations compared to the $\Lambda$CDM model; this can lead to a fast mode in the DM perturbation evolution~\cite{Weinberg:2002kg,Voruz:2013vqa} and thus to a gravitational coupling between DM and photons that suppresses the odd (compression) CMB peaks.
\end{itemize}

Since we want to derive Lyman-$\alpha$ bounds on DM-DR models, we are interested in the behaviour of the matter power spectrum $P(k)$ on small scales. As described in Refs.~\cite{Buckley:2014hja,Archidiacono:2017slj}, the effect of the coupling between DM and DR is twofold:
\begin{itemize}
\item The late kinetic decoupling induced by DM-DR scattering yields a collisional damping of the matter power spectrum. The damping translates into a cut-off in the halo mass function, thus providing the solution to the missing satellite problem.
\item Besides the exponential damping (only apparently similar to warm DM), the opposite forces of DM gravitational clustering and DR relativistic pressure may lead to a series of so-called Dark Acoustic Oscillations (DAO) typical of models of DM-DR interactions mediated by a new light mediator~\cite{Cyr-Racine:2013fsa}.
\end{itemize}

A special comment has to be dedicated to a class of models discussed in Refs~\cite{Buen-Abad:2015ova, Lesgourgues:2015wza, Buen-Abad:2017gxg}, like for instance Non-Abelian Dark Matter (NADM), in which the momentum transfer rate from DM to DR, related to the ETHOS rate by $\Gamma = - \Gamma_\mathrm{DM-DR}/a$, scales like $a^{-2}$. In this case the suppression of the matter power spectrum is smooth, as the temperature dependence of the interaction rate ($n=0$ in the ETHOS parametrisation) is the same as the temperature dependence of the expansion rate during the radiation dominated epoch. Moreover, DR particles tend to have strong self-interactions caused either by their charge under the new gauge group of the dark sector, or by the fact that they are the gauge bosons of this group. These models are described by the parameters:
\begin{itemize}
\item $\Delta N_\mathrm{fluid} \equiv \frac{\rho_\mathrm{dr}}{\rho_{1\nu}}$, which gives the amount of self-interacting DR, parametrised as the effective number of extra neutrino families, and
\item $\Gamma_0 \equiv \Gamma \, (a/a_0)^2$, which gives the momentum transfer rate from dark matter to DR at redshift $z = 0$.
\end{itemize}
Thus they can be described with the ETHOS parametrisation in the $n=0$ case, provided that DR is treated as a perfect fluid.

%% file: Methods.tex
\section{Methods}
\label{sec:methods}
\subsection{Lyman-$\alpha$ likelihood}

\begin{figure}[h!]
\centering
\begin{tabular}{cc}
\includegraphics[width=0.45\linewidth]{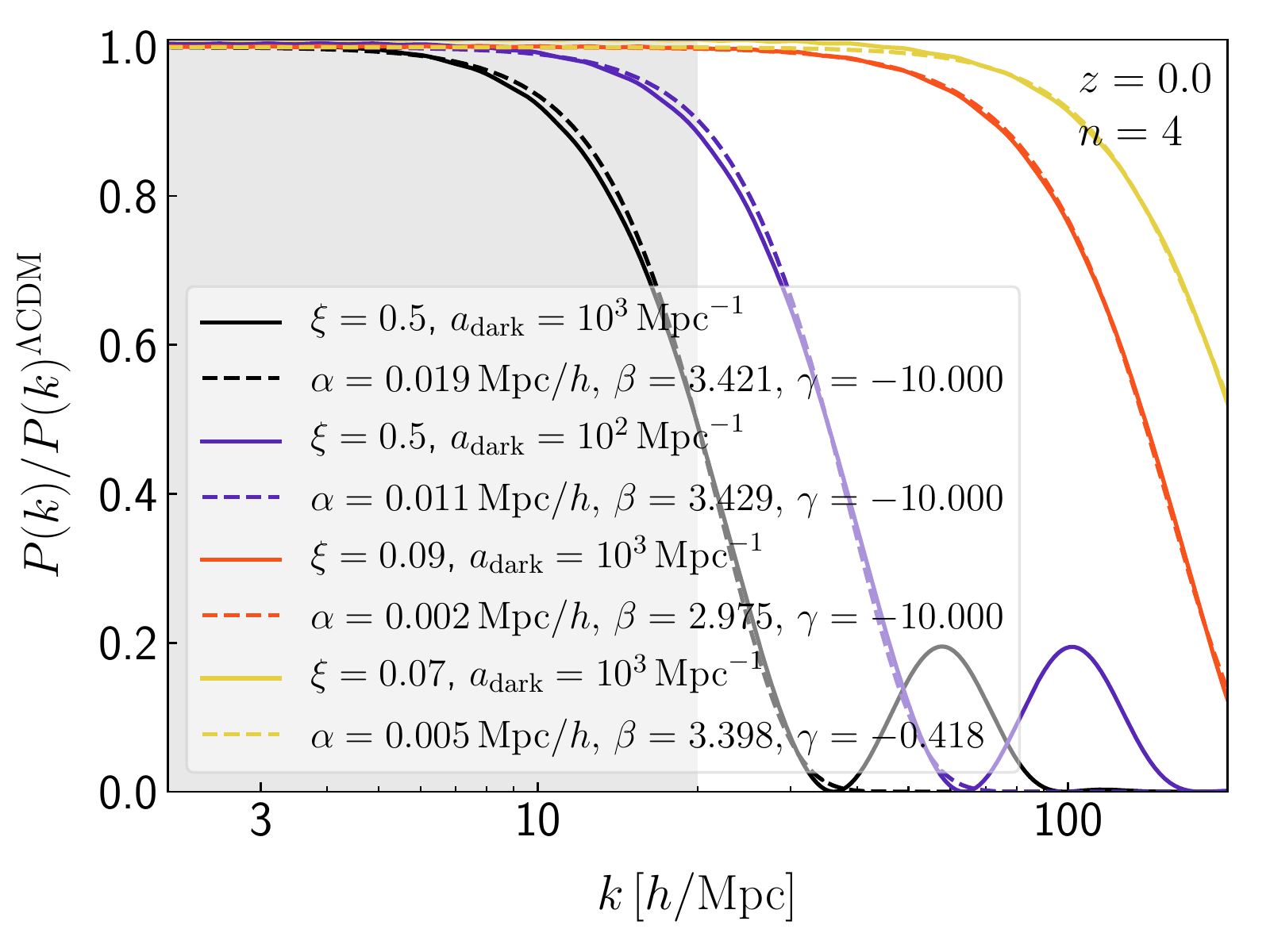}
&\includegraphics[width=0.45\linewidth]{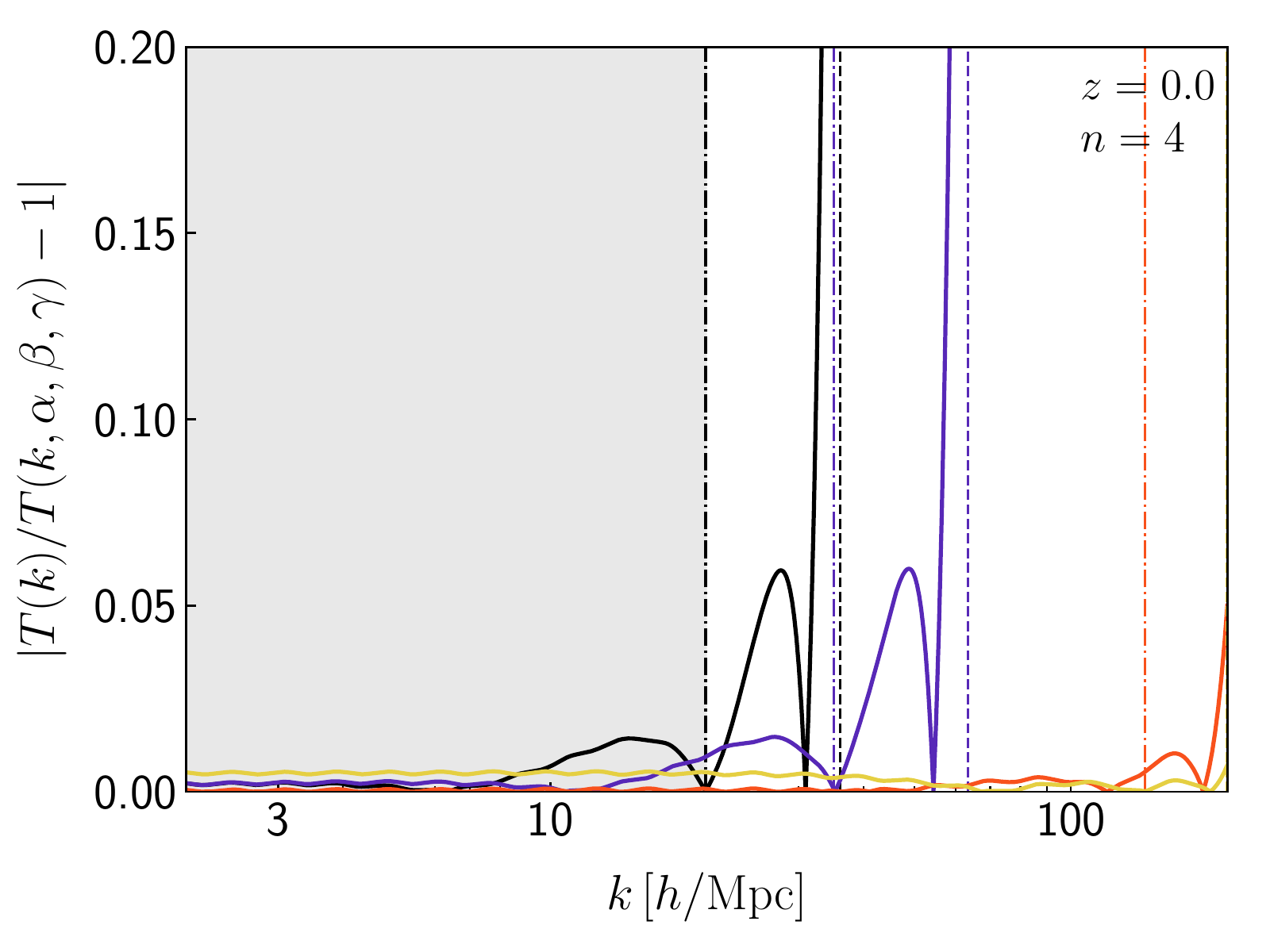}\\
\includegraphics[width=0.45\linewidth]{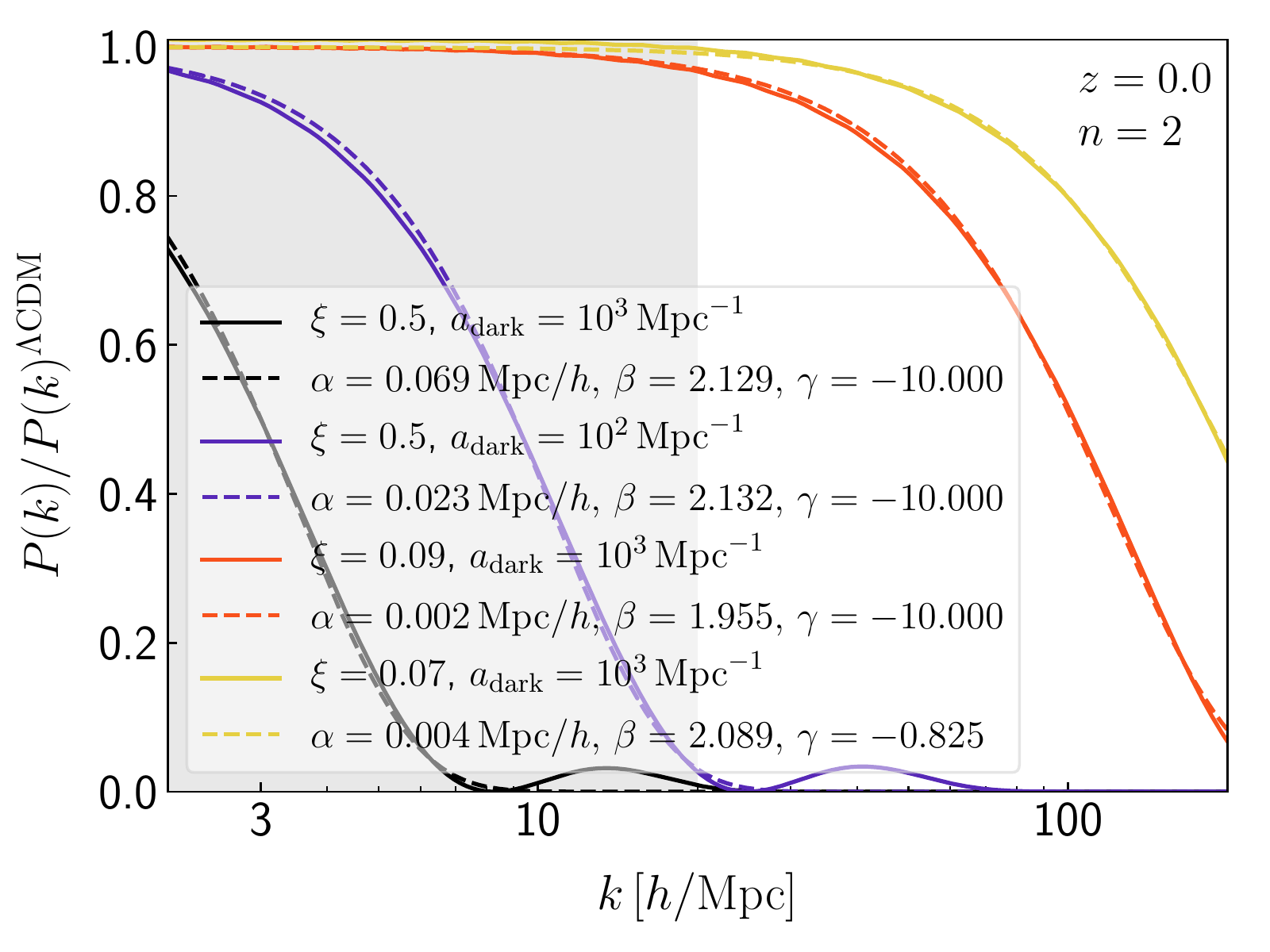}
&\includegraphics[width=0.45\linewidth]{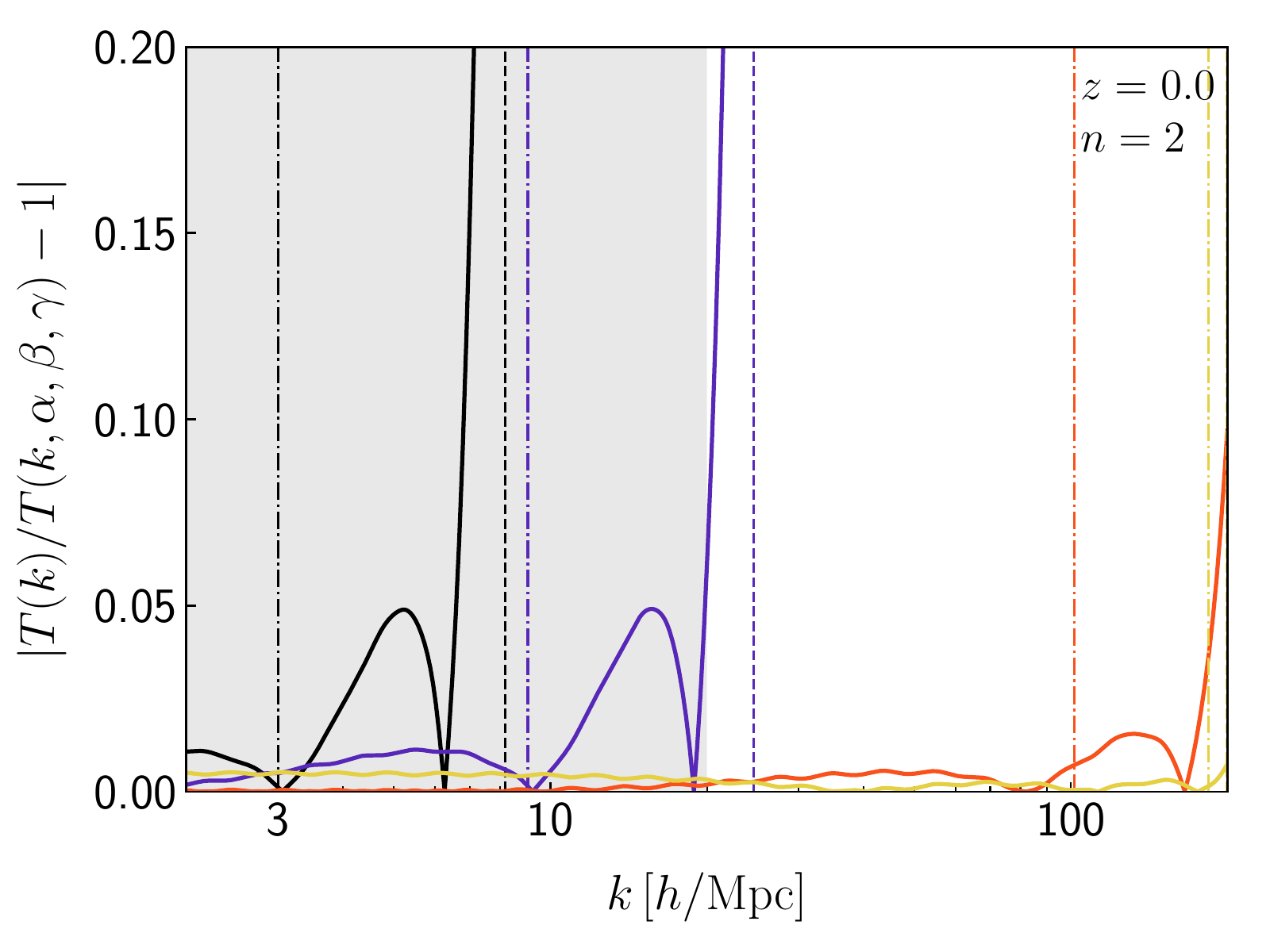}\\
\includegraphics[width=0.45\linewidth]{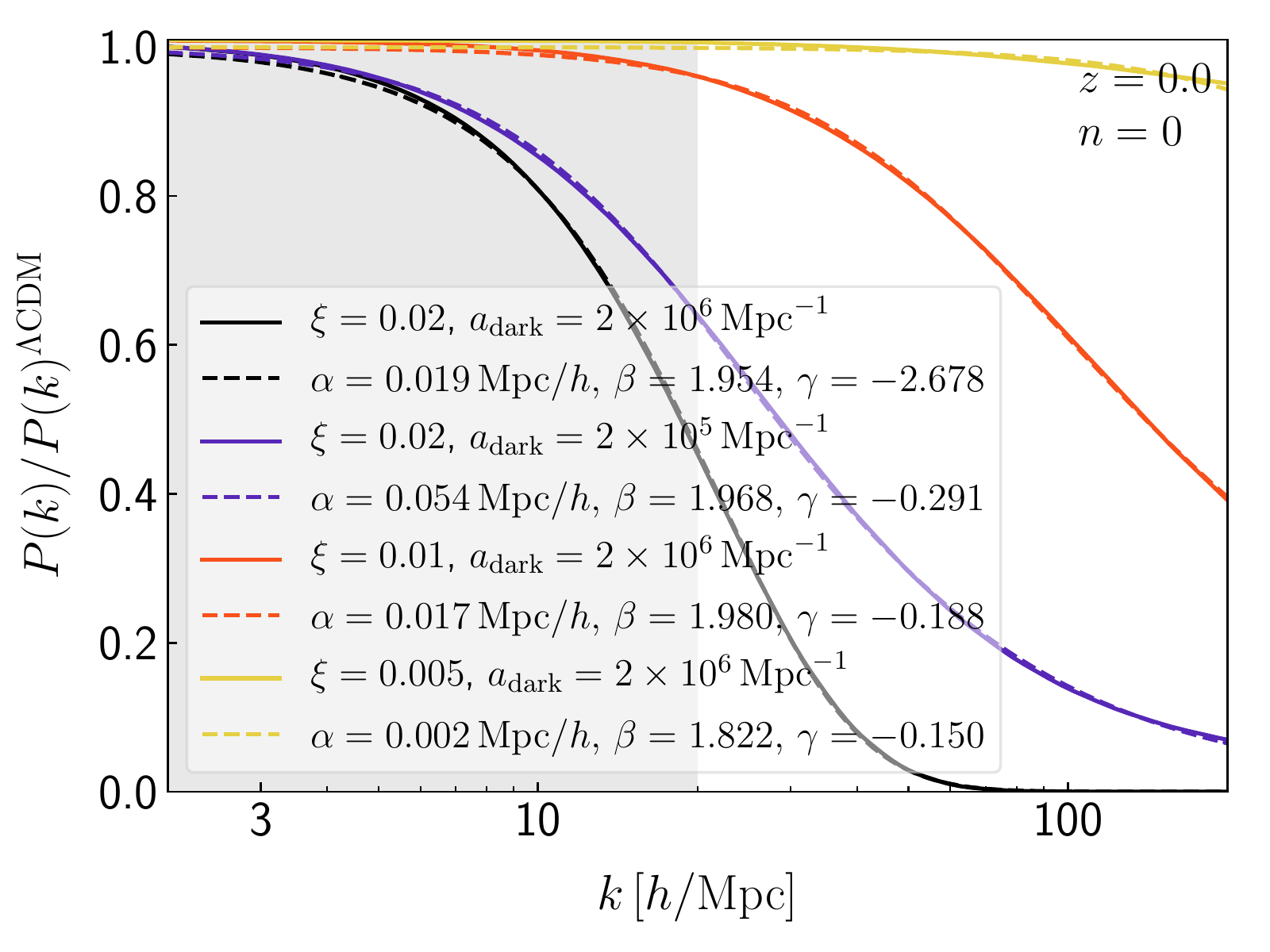}
&\includegraphics[width=0.45\linewidth]{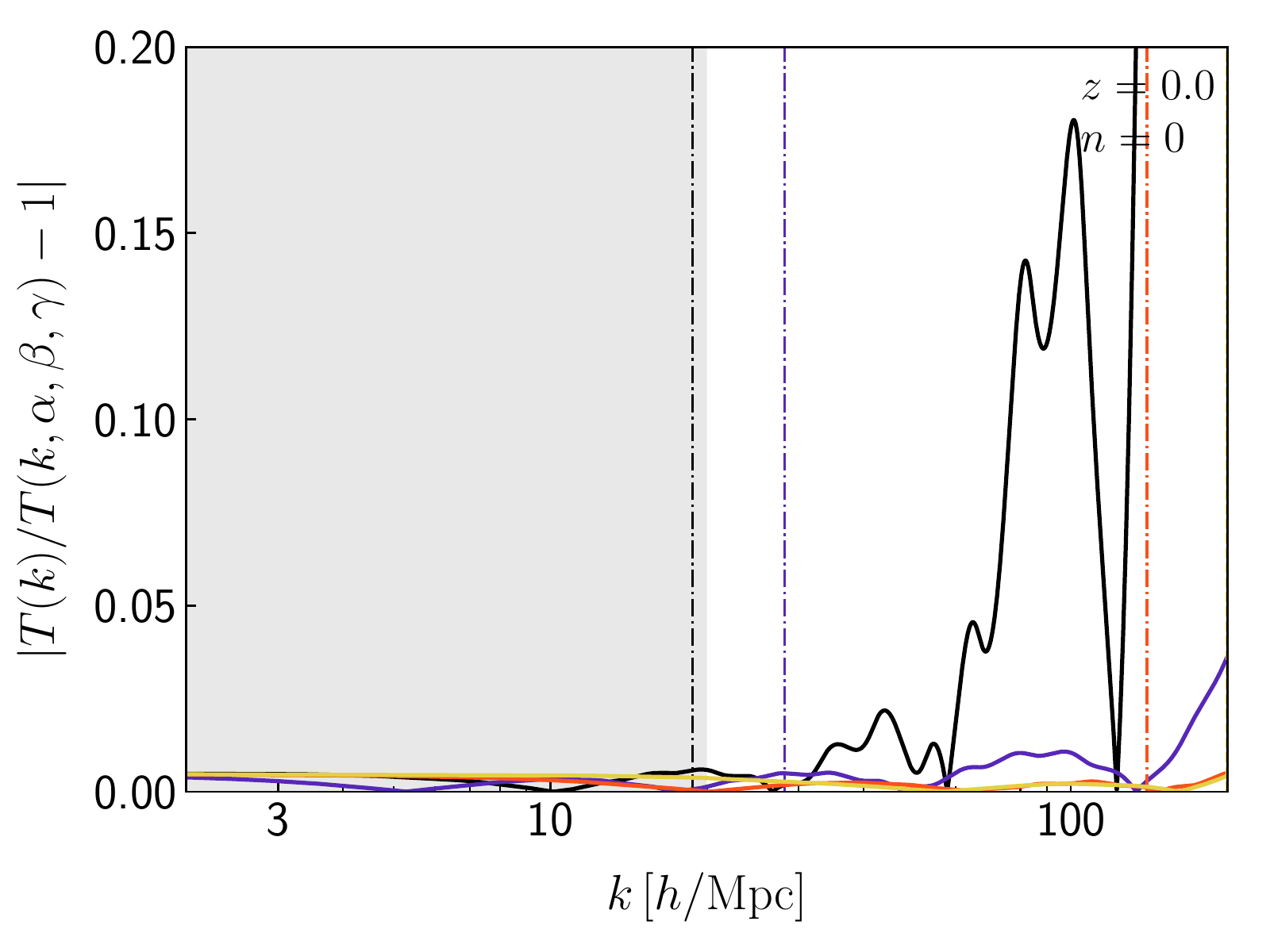}\\
\end{tabular}
\caption{
\emph{(Left)} Linear transfer functions $T(k)^2=P(k)/P(k)^{\Lambda \rr{CDM}}$ at $z=0$, for $n=4$ (top row), $n=2$ (second row), $n=0$ (bottom row). The different colours correspond to different values of the amount of dark radiation $\xi$ and of the strength of the interaction $a_\rr{dark}$. Solid lines depict the true $T(k)^2$, while dashed lines of the same colour show the corresponding $\{\alpha, \beta, \gamma \}$-fit. \emph{(Right)} Relative deviation of the $\{\alpha, \beta, \gamma \}$-fit from the true $T(k)^2$ (solid lines) for the same models (colours) as in the left panel. The vertical lines show 
$k_{1/2}$ (dot-dashed lines) and $k_\rr{fit}$ (dashed lines - for $n=0$, $k_\rr{fit}=k_\rr{max}$). The grey shaded region approximately represents the $k$ range probed by Lyman-$\alpha$ data.
}
\label{fig:Tk_lin}
\end{figure}
\begin{figure}[h!]
\centering
\begin{tabular}{cc}
\includegraphics[width=0.45\linewidth]{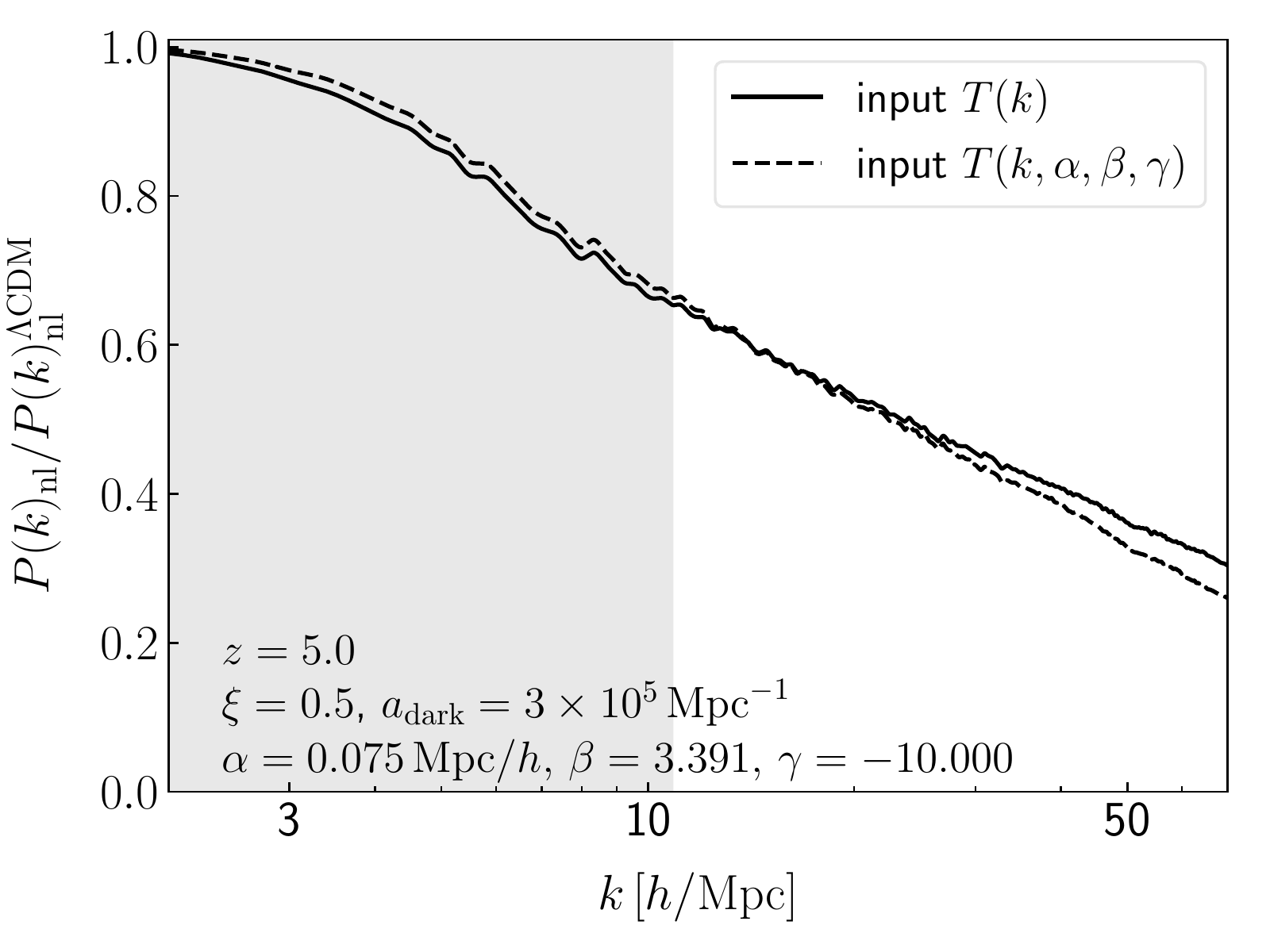}
&\includegraphics[width=0.45\linewidth]{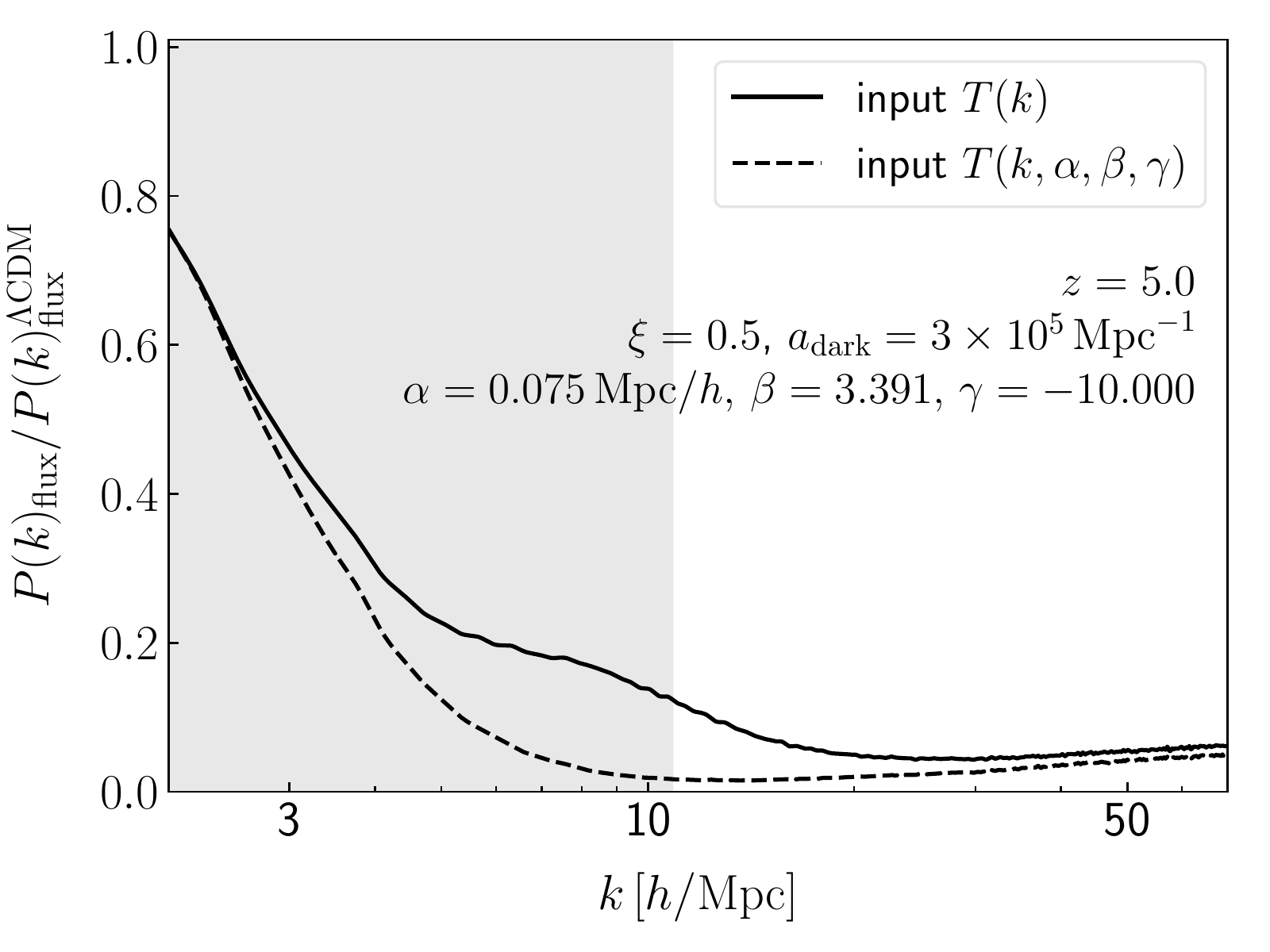}
\end{tabular}
\caption{
Ratio between the non-linear matter power spectra (left panel) and the corresponding ratio of 1D flux power spectrum (right panel) at $z=5$.
The spectra are obtained from simulations with the linear input given either by the true $T(k)$ (solid lines) or by the fit $T(k,\alpha,\beta,\gamma)$. The theoretical model is $n=4$ and it has $\xi=0.5$ and $a_\rr{dark}=3\times10^5\,\rr{Mpc}^{-1}$.
The grey shaded region defines the $k$ range of MIKE/HIRES data.
}
\label{fig:Tk_nl}
\end{figure}

In order to provide limits on the properties of interactiong DM-DR scenarios from the Lyman-$\alpha$ forest, we have devised a new {\sc MontePython}~\cite{Audren:2012vy,Brinckmann:2018cvx}  
likelihood, based on the general parametrisation introduced in Ref.~\cite{Murgia:2017lwo}. The corresponding data set is the HIRES/MIKE samples of quasar spectra, which were obtained with the HIRES/KECK spectrograph and the MIKE/Magellan spectrograph, at redshift bins $z=4.2,4.6,5.0,5.4$, in 10 $k$-bins in the interval $0.001-0.08$ s/km, with spectral resolution of $6.7$ and $13.6$ km/s, for HIRES and MIKE, respectively~\cite{Viel:2013apy}. As in the analyses of Refs.~\cite{Viel:2013apy,Irsic:2017ixq, Murgia:2018now}, we applied a conservative cut on the flux power spectra, by using only the measurements at $k > 0.005$ s/km, in order to avoid large-scale systematic uncertainties due to continuum fitting. 
Moreover, we did not include in our analyses the highest redshift bin for MIKE data, for which the errors on the flux power spectra are very large (see Ref.~\cite{Viel:2013apy} for more details). We have thereby used $49$ $(k,z)$ data points.

The new likelihood takes advantage of the scheme described in Ref.~\cite{Murgia:2018now}, which allows to interpolate between different cosmological models without the need of running dedicated numerical simulations. Such procedure relies in fact on a large set of pre-computed hydrodynamical simulations, and on an advanced interpolation method which is able to accurately deal with the sparse, non-regular grid defined by the simulations.
As in Refs.~\cite{Irsic:2017ixq, Murgia:2018now}, our reference model simulation has a box length of $20/h$ comoving Mpc with $2 \times 768^3$ gas and CDM particles (with gravitational softening $1.04/h$ comoving kpc) in a flat $\Lambda$CDM universe
with cosmological parameters $\Omega_m = 0.301$, $\Omega_b = 0.0457$, $n_s = 0.961$, $H_0 = 70.2$~km~s$^{-1}$~Mpc$^{-1}$ and $\sigma_8 = 0.829$~\cite{Ade:2015xua}.

The flux power spectrum is affected both by astrophysical and cosmological parameters. The latter ones are poorly constrained by Lyman-$\alpha$ forest data alone~\cite{Murgia:2018now}, which is why the possibility of combining them with CMB data from Planck constitutes a big advantage with respect to previous analyses. The astrophysical parameters impact our likelihood as nuisance parameters, to marginalise over. Our multidimensional grid has been built in order to include several values of the aforementioned parameters, spanning all the viable volume of the corresponding parameter space.

Concerning the cosmological parameters, we focused on $\sigma_8$, i.e. the normalisation of the linear matter power spectrum, and $n_{\rm eff}$, namely the slope of the matter power spectrum at the scale of the Lyman-$\alpha$ forest ($0.009$ s/km), given that varying these two parameters is sufficient to properly accounting for the effect on the matter power spectrum due to changes in its initial slope and amplitude~\cite{seljak2006,McDonald:2004eu,Arinyo-i-Prats:2015vqa}. We thus considered five different simulations for both $\sigma_8$ (in the range $[0.754, 0.904]$) and $n_{\rm eff}$ (in the interval $[-2.3474,-2.2674]$).  Additionally, we included three simulations corresponding to different values for the instantaneous reionization redshift,~i.e.~$z_{\rm reio} = 7,9,15$, with $z_{\rm reio} = 9$ being our reference value.

Concerning the astrophysical parameters, we modelled the thermal history of the IGM in the form of the
amplitude $T_0$ and the slope $\widetilde{\gamma}$ of its temperature-density
relation, parametrised as $T=T_0(1+\delta_\rr{IGM})^{\widetilde{\gamma}-1}$, with $\delta_{\rr{IGM}}$ being the IGM overdensity~\cite{hui97}.
The corresponding grid points are given by three different simulations with temperatures at mean density $T_0(z = 4.2) = 6000, 9200, 12600$ K,
evolving with redshift, as well as a set of three values for the slope of the temperature-density relation, $\widetilde{\gamma}(z = 4.2) = 0.88, 1.24, 1.47$.
The redshift evolution of both $T_0$ and $\widetilde{\gamma}$ are parametrised as power laws, such that $T_0(z) = T_0^A[(1+z)/(1+z_p)]^{T_0^S}$ and $\widetilde{\gamma}(z) = \widetilde{\gamma}^A[(1+z)/(1+z_p)]^{\widetilde{\gamma}^S}$, where the pivot redshift $z_p$ is the redshift at which most of the Lyman-$\alpha$ forest pixels are coming from ($z_p = 4.5$ for MIKE/HIRES).
The reference thermal history is defined by $T_0(z = 4.2) = 9200$ and $\widetilde{\gamma}(z = 4.2) = 1.47$, since such values provide a good fit to observations~\cite{bolton17}.
Furthermore, we considered the effect of ultraviolet (UV) fluctuations of the ionizing background, the impact of which is encoded in the parameter $f_{\rm UV}$. The corresponding template is built from a set of three simulations with $f_{\rm UV} = 0, 0.5, 1$, where
$f_{\rm UV} = 0$ corresponds to a spatially uniform UV background~\cite{Irsic:2017ixq}.
Finally, we have several grid points associated to different values for the mean Lyman-$\alpha$ forest flux $\bar{F}(z)$, obtained by
selecting 9 different values for it, namely $(0.6,0.7,0.8,0.9,1.0,1.1,1.2,1.3,1.4) \times \bar{F}_{\rm REF}$, with the reference values being the ones of the SDSS-III/BOSS measurements~\cite{boss2013}. In order to have a more refined grid in terms of mean fluxes, we also included 8 additional values, obtained by rescaling the optical depth $\tau = -\ln\bar{F}$,~i.e.~$(0.6,0.7,0.8,0.9,1.1,1.2,1.3,1.4) \times \tau_{\rm REF}$.

According to Refs.~\cite{Murgia:2017lwo, Murgia:2017cvj, Murgia:2018now}, the non-standard DM sector of the parameter space is described through a set of three parameters,  dubbed as $\alpha$, $\beta$, and $\gamma$, which are associated to the scale and shape of the power suppression with respect to $\Lambda$CDM induced by the DM-DR interaction. Roughly speaking, $\alpha$ specifies the scale of the cut-off, and is related to the value $k_{1/2}$ where the spectrum is suppressed by 50\% compared to the CDM case;  the shape of the step for $k<k_{1/2}$ depends mainly on $\beta$, while that for $k>k_{1/2}$ -- i.e. the shape of the tail-- depends on both $\beta$ and $\gamma$ (see Refs.~\cite{Murgia:2017lwo,Murgia:2017cvj} for details). We used indeed a $\{\alpha, \beta, \gamma \}$-grid constituted by 109 hydrodynamical simulations (512$^3$ particles in a 20 Mpc$/h$ box), obtained keeping the astrophysical and cosmological parameters fixed to their reference values.
Additionally, we also included 8 hydrodynamical simulations, in which the values for $\alpha$ correspond to thermal WDM masses of 2,3,4,5,6,7,8,9~keV, $\beta$ and $\gamma$ are fixed to their thermal values, and all the other cosmological and astrophysical parameters are fixed to their reference values. The full set of simulations consists thereby in 117 points thoroughly sampling the volume of the $\{\alpha, \beta, \gamma \}$-space~\cite{Murgia:2018now}.

All simulations were run with the hydrodynamic N-body code {\sc gadget-3}, a modified version of the publicly available numerical code {\sc gadget-2}~\cite{Springel:2000qu,Springel:2005mi}. The initial conditions were produced by displacing the DM particles from a cubic Cartesian grid according to second-order Lagrangian Perturbation Theory, with the {\sc 2LPTic} code~\cite{Crocce:2006ve}, at redshift $z=99$. Let us remark that the non-standard nature of the dark sector is accurately followed only at the linear level,~i.e.~its impact is assumed to be fully encoded in the suppressed initial power spectra produced by our modified version of {\sc class} and used as inputs for {\sc 2LPTic}. While during the non-linear structure evolution investigated by our numerical simulations, DM is treated as standard, pressureless CDM. The motivation for this treatment is twofold. First, DM-DR interactions have significant effects only at earlier times ($z>99$) 
\cite{Cyr-Racine:2015ihg}. Second, DM self-interactions -- which are expected to be relevant at late times -- can also be safely neglected during the non-linear evolution, since the scales probed by Lyman-$\alpha$ are somewhat too large to be affected by such exotic DM properties~\cite{Vogelsberger:2015gpr} (see,~e.g.~Ref.~\cite{Nori:2018pka}, where this is explicitly demonstrated in the analogous context of small-scale power suppression induced by ultra-light scalar DM).

The interpolation is done in terms of ratios between the flux power spectra of the non-standard DM models and the reference $\Lambda$CDM one, by using the so-called \emph{Ordinary Kriging} method~\cite{webster2007geostatistics}. We first interpolate in the astrophysical and cosmological parameter space for the $\Lambda$CDM case,~i.e.~in the $\alpha=0$ plane. We then correct all the $\{\alpha, \beta, \gamma \}$-grid points accordingly, and we finally interpolate in the $\{\alpha, \beta, \gamma \}$-space. This procedure relies on the assumption that the corrections due to non-reference astrophysical or cosmological parameters are universal,~i.e.~we can apply the same corrections computed for the $\Lambda$CDM case ($\alpha=0$) to all the non-standard DM models described by our parametrisation. The robustness of such procedure has been extensively discussed in Ref.~\cite{Murgia:2018now}~(see also Ref.~\cite{Murgia:2019duy}, where a similar approach has been used to test Primordial Black Hole scenarios).

The main advantage of the $\{\alpha, \beta, \gamma \}$-parametrisation is that it allows to systematically explore the parameter space of any non-standard DM cosmological model, provided that the corresponding linear power spectrum can be fitted in terms of the three aforementioned parameters.
The new likelihood directly translates the limits on $\alpha$, $\beta$, and $\gamma$ obtained through MIKE/HIRES data into constraints on the fundamental particle physics parameters.
The scheme is the following:
\begin{itemize}
 \item The linear matter power spectrum associated to a given combination of cosmological parameters (six $\Lambda$CDM parameters plus additional non-standard DM parameters) is produced by {\sc class} up to a maximum wavenumber chosen to be $k_\rr{max}=200\,h/\rr{Mpc}$.  Corresponding values of the derived parameters $(\sigma_8, n_\rr{eff}, z_\rr{reio})$ used to define the N-body simulations are computed.
When these values fall outside of the conservative range assumed in the simulations,  the model can safely be rejected, given that such models would be very bad fits to the Planck data (this will be further cross-checked in some dedicated runs called ``Lyman-$\alpha$ prior'', discussed in section~\ref{sec:results}). The only exception is the case of models with a low reionisation redshift. The prior used in the N-body grid, $7<z_\rr{reio}<15$, was motivated by Planck 2013 results. Instead Planck 2015 + BAO data are compatible with $z_\rr{reio}=8.7\pm1.1$ (68\%CL), such that in our runs, models with $6<z_\rr{reio}<7$ might still be acceptable fits and should not be systematically rejected. In practice, within our Lyman-$\alpha$ likelihood, we re-map any $6<z_\rr{reio}<7$ to $z_\rr{reio}=7$. This is a satisfactory approximation given that the value of the reionization redshift has a small impact on the flux power spectrum.
 \item The linear matter power spectrum of the ``equivalent'' $\Lambda$CDM model is also produced. Note that in Ref.~\cite{Murgia:2018now}, the grid of N-body simulations for $\Lambda$CDM models assumes a fixed standard value of the ultra-relativistic relic density, corresponding to $N_\rr{eff}=3.046$. In general, for models of warm or interacting DM with the same $N_\rr{eff}$, computing the spectrum of the ``equivalent'' $\Lambda$CDM model  would be very straightforward: we would just need to re-run {\sc class} with an infinite DM mass and/or zero interaction rates. However, in the present paper, all models include DR and an enhanced value of $N_\rr{eff}$. To deal with this, we use the accurate procedure described in Ref.~\cite{Rossi:2014nea}, which allows to re-map a $\Lambda$CDM model with $N_\rr{eff}>3.046$ to another one sharing the same matter power spectrum up to some scale, but with $N_\rr{eff}=3.046$:  this can be achieved by adjusting the value of other cosmological parameters according to some analytic relations. In other words, for each model with $N_\rr{eff}>3.046$, our Lyman-$\alpha$ likelihood automatically reformulates the problem in terms of an equivalent $\Lambda$CDM model with $N_\rr{eff}=3.046$, for which we study the effect of a suppression in the small-scale matter power spectrum caused only by non-standard DM effects.
  \item The transfer function, i.e. the square root of the ratio between the two power spectra is fitted in terms
 of $\{\alpha, \beta, \gamma \}$ with a simple least squared method. The fitting algorithm only includes points until a finite value $k_\rr{fit}$ which is set by default to $k_\rr{max}=200\,h/\rr{Mpc}$. However, for transfer functions with oscillations within the range $[0, k_\rr{max}]$, $k_\rr{fit}$ is reduced to the first zero of the function. The fit is also restricted to values of $\{\alpha, \beta, \gamma \}$ within the region covered by the grid of simulations: $0 \leq \alpha \leq 0.17$, $1.5 \leq \beta \leq 10$, and $-10 \leq \gamma \leq -0.15$. Furthermore, if the difference between the ``exact'' transfer function and the fitted one is too large in a region in which the power spectrum is not strongly suppressed, our method cannot be considered accurate and reliable enough. Thus we need to implement a conservative ``applicability check'' rejecting models giving bad $\{ \alpha, \beta, \gamma\}$-fits, but such that the Lyman-$\alpha$ data remain more constraining than the applicability check itself. If this condition is met, this check is just a technical step, not biasing our final results, because rejected models would anyway conflict the data. In practice, our likelihood requires that the $\{ \alpha, \beta, \gamma\}$-fit to the transfer function is accurate to better than 10\% in the whole region where this function is larger than $0.2$. The 10\% accuracy is sufficient for data with statistical uncertainties of $\sim$ 10\% such as in Ref. \cite{Irsic:2017sop}. In section~\ref{sec:results}, we will describe some dedicated runs proving that this applicability check is much less constraining than the data and has no impact on our final results.
 \item At this point, if the considered model has passed the aforementioned applicability checks, its flux power spectrum is produced by performing the interpolation procedure described above. By confronting such flux power spectrum against Lyman$-\alpha$ forest data, a $\chi^2$-value is associated to the corresponding combination of parameters.
 \item The procedure is iterated per each MCMC step, until convergence is reached, i.e. until accurate constraints on the cosmological and astrophysical parameters of the model are determined.
 \item At each step, the fitted values of $\{\alpha, \beta, \gamma \}$ are kept in the MCMC chains as derived parameters, to check a posteriori the range of power spectrum shapes covered by a given cosmological model.
\end{itemize}

In order to show how the pipeline described above works, we plot in the left panels of Fig. \ref{fig:Tk_lin} the square of the linear transfer function of a few selected DM-DR models, or in other words, their linear matter power spectrum divided by that of the  $\Lambda$CDM equivalent model. By construction, the transfer function always has an asymptote of one in the small-$k$ limit. For each model, we compare it with its best fit using the $\{\alpha, \beta, \gamma \}$-parametrisation. In the right panels we show the relative error of the fit.
Notice that the $\{\alpha, \beta, \gamma \}$ cannot reproduce the oscillations in $T(k)$ after the first zero (for $k>k_\rr{fit}$). However, the power of the subsequent oscillations is small. Fig. \ref{fig:Tk_nl} demonstrates that the impact of these oscillations on structure formation is negligible and located at scales smaller than those probed by Lyman-$\alpha$, even for a rather large interaction strength.
As it has already been shown in Ref.~\cite{Murgia:2018now}, significant differences between the ``exact'' flux power spectrum and the $\{\alpha,\beta,\gamma\}$-prediction appear only when the power suppression with respect to the standard CDM case is more than 50$\%$,~i.e.~for models whose power spectra lie very far from the Lyman-$\alpha$ forest data points. This fully justifies ignoring such oscillations when applying our fit (see Ref.~\cite{Murgia:2018now} for a detailed discussion).
For $k \lesssim k_\rr{fit}$, the $\{\alpha, \beta, \gamma \}$-parametrisation works rather well in reproducing the real transfer function (see Fig. \ref{fig:Tk_lin}). The relative error features a small bump at scales $k_{1/2}<k<k_\rr{fit}$, with an amplitude related to the DR content. 
Then it diverges at $k \longrightarrow k_\rr{fit}$, which is unavoidable given that the reference transfer function goes to zero, but harmless given the small power of those scales. This is not a problem for our applicability check, which only applies up to the wavenumber at which the transfer function crosses 0.2.

We stress one important point here. The Lyman-$\alpha$ forest likelihood built in this work significantly improves over previous likelihood analyses present in the literature addressing DM-DR interactions (e.g. \cite{Krall:2017xcw}). In previous works, the likelihood was based on an estimate of the linear matter power spectrum amplitude slope and curvature obtained from low resolution and low signal-to-noise SDSS-II data \cite{McDonald:2004eu}. However, these measurements were derived only in the standard $\Lambda$CDM model and are expected to be valid only for small deviations around this model. There exists no explicit proof that models with interacting DM-DR fall inside the range of validity of this method.
In this work, we fit instead the observed quantity, the 1D flux power, using a set of dedicated simulations that take in input the linear power spectra of the $\{\alpha, \beta, \gamma \}$-parametrisation, thus fully taking into account the cosmological signature of DM-DR interactions. A dedicated investigation of the non-linear evolution of structures in interacting DM-DR scenarios using N-body/hydro simulations has only been performed in this work and in Ref.~\cite{Bose:2018juc}. The focus of the latter reference was on the survival of oscillatory features in the flux power spectrum, and not on a full MCMC analysis of the flux power.
Furthermore, the data set used in this work is expected to be the most updated and constraining one for models with a small-scale suppression. Indeed, high resolution high signal-to-noise quasar spectra can go down to the smallest scales probed by IGM structures \cite{Viel:2013apy,Boera:2018vzq}. Low resolution data from surveys like SDSS have smaller statistical errors but are limited to larger scales. Thus they are more appropriate for constraining neutrino masses and/or cosmological parameters \cite{boss2013}, but less constraining for small-scale features.

\subsection{DM-DR interactions in {\sc class}}
\label{sec:class}

{\sc class} already incorporated several DM species and related input parameters: the CDM sector (including the effects of energy release from DM annihilation or decay into electromagnetic particles~\cite{Poulin:2016anj}); the decaying DM sector assuming a decay into DR~\cite{Poulin:2016nat}; and the non-cold DM sector featuring an arbitrary number of non-cold species, covering most warm DM models~\cite{Lesgourgues:2011rh}.

The ETHOS framework for an effective description of DM-DR interactions~\cite{Cyr-Racine:2015ihg} was already implemented in {\sc class} by some of us in Ref.~\cite{Archidiacono:2017slj}, as a set of modifications to the existing CDM equations. For the purpose of this work (and for the public release of the code that will follow), we re-implemented the same set of equations in {\sc class}, but for a new ``interacting DM species'' coexisting with the plain CDM species and enlarging the total number of DM sectors in {\sc class}. This allows to investigate mixed DM models, and it has an appropriate structure for accommodating in future versions more types of DM interactions (e.g. DM-baryon or DM-photon), either separately or at the same time. For the moment, the new sector includes parameters like the DM mass  ({\tt m\_dm}), the fraction of the total CDM density ({\tt f\_idm\_dr}) , as well as other parameters related to the ETHOS model, fully described in {\tt explanatory.ini}, and appearing here in typefaces.

Here we only recall the main equations of DM and DR perturbations in the ETHOS model, and we refer the reader to Refs.~\cite{Archidiacono:2017slj,Cyr-Racine:2015ihg} for details. 

As already mentioned in section~\ref{sec:theory}, the amount of DR is set by the temperature ratio $\xi=T_\rr{DR}/T_\gamma$ ({\tt xi\_idr}), and its physical density is:
\begin{equation}
\omega_\rr{DR}= \left(\frac{g_\rr{DR}}{2}\right) f_\rr{DR} \xi^4 \omega_\gamma,
\label{eq:omega_DR}
\end{equation}
where the statistical factor $f_\rr{DR}$ ({\tt stat\_f\_idr}) is $7/8$ for fermionic DR and $1$ for bosonic,
$g_\rr{DR}$ is the DR number of internal degrees of freedom and it is assumed to be $2$.
The input parameter {\tt idr\_nature} describes the DR nature, i.e. free-streaming or fluid:
In the former case the DR hierachy is evolved up to $\ell_\rr{dark}$ ({\tt l\_max\_idr}) (set by default to 17),
while in the latter only the modified continuity and Euler equations are present.
In the free-streaming case, the DR hierarchy in Newtonian gauge is:
\begin{gather}
\dot{\delta}_\rr{DR}+\frac{4}{3}\theta_\rr{DR}-4\dot{\phi}=0,\\
\dot{\theta}_\rr{DR}+k^2\left(\sigma_\rr{DR}-\frac{1}{4}\delta_\rr{DR}\right)-k^2 \psi=\nonumber \\
\Gamma_\rr{DR-DM}\left(\theta_\rr{DR}-\theta_\rr{DM}\right), \\
\dot{\pi}_{\rr{DR},\ell}+\frac{k}{2\ell+1}\left((\ell+1)\pi_{\rr{DR},\ell+1}-\ell\pi_{\rr{DR},\ell-1}\right)=\nonumber \\
\left( \alpha_\ell \Gamma_\rr{DR-DM}+\beta_\ell \Gamma_\rr{DR-DR}\right) \pi_{\rr{DR},\ell}, \,\,\, 2 \leq \ell \leq \ell_\rr{dark}.
\label{eq:dr2}
\end{gather}
The density and velocity dispersion perturbations are labelled as $\delta$ and $\theta$, respectively,
the DR shear perturbation is $\pi_\rr{DR}=2\sigma_\rr{DR}$,
 $\phi$ and $\psi$ are the gravitational potentials.
The specifications related to the DR-DM interactions are embedded into
$\Gamma_\rr{DR-DM}$, which is the comoving interaction rate (see the formula and the discussion below),
and $\alpha_\ell$ ({\tt alpha\_dark}) is the array of the interaction angular coefficients for $\ell=2, ..., \ell_\rr{dark}$.
The DR self-interactions are encoded in the comoving rate $\Gamma_\rr{DR-DR}$, whose strength is {\tt b\_dark}, and whose angular coefficients are $\beta_\ell$ ({\tt beta\_dark}).
The DM perturbation equations are:
\begin{gather}
\dot{\delta}_\rr{DM}+\theta_\rr{DM}-3\dot{\phi}=0,\\
\dot{\theta}_\rr{DM}-k^2c_\rr{DM}^2\delta_\rr{DM}+ \mathcal{H} \theta_\rr{DM}-k^2\psi=\nonumber \\
\Gamma_\rr{DM-DR} \left( \theta_\rr{DM} - \theta_\rr{DR}\right),
\end{gather}
where $c^2_\rr{DM}$ is the dark sound speed.
The interactions are embedded in the right-hand-side of DR and DM dipole equations and of DR higher order momenta. The effective comoving scattering rate of DR off DM can be parametrised as:
\begin{equation}
\Gamma_\rr{DR-DM}=-\Omega_\rr{DM}h^2 a_\rr{dark} \left(\frac{1+z}{1+z_d} \right)^n,
\label{eq:DRDM}
\end{equation}
where $z_d=10^7$ is a normalization factor,
$n$ ({\tt nindex\_dark}) is the temperature dependence,
and $a_\rr{dark}$ ({\tt a\_dark}) is the interaction strength.
Applying energy-momentum conservation, we obtain: 
\begin{align}
\Gamma_\rr{DM-DR}&=R_\rr{dark}\Gamma_\rr{DR-DM} \nonumber \\
&=\left(\frac{4}{3} \frac{\rho_\rr{DR}}{\rho_\rr{DM}} \right)\Gamma_\rr{DR-DM}.
\label{eq:DMDR}
\end{align}

With respect to Ref.~\cite{Archidiacono:2017slj}, the present version of the code implements the tight-coupling regime between DM and DR.  By default, {\sc class} uses a stiff integrator ({\tt ndf15}) \cite{Blas:2011rf} for the perturbation equations, which means that rather large values of the interaction rate can be reached while using the default equations and keeping the code fast. However, in order to investigate the very small scales probed by Lyman-$\alpha$, the tight-coupling is required. This regime is fully operational in our released {\sc class} version.
The tightly-coupled equations are switched on automatically whenever the ratio
between the conformal interaction rate and Hubble times, $\mathcal{H}/\Gamma_\rr{DR-DM}$, falls below a threshold set by default to 0.005, and the ratio between the conformal interaction and acoustic oscillation times, $k/\Gamma_\rr{DR-DM}$, falls below 0.01\footnote{The two thresholds are defined as the precision parameters {\tt dark\_tight\_coupling\_trigger\_tau\_c\_over\_tau\_h} and {\tt dark\_tight\_coupling\_trigger\_tau\_c\_over\_tau\_k}.}.
At first order in $\Gamma_\rr{DR-DM}^{-1}$ the DM-DR slip is:
\begin{align*}
\dot{\Theta}^\rr{TCA}_\rr{DM-DR}&=\dot{\theta}_\rr{DR}-\dot{\theta}_\rr{DM}\\
&=\left(n-\frac{2}{1+R_\rr{dark}}\right)\frac{\dot{a}}{a} \left(\theta_\rr{DM}-\theta_\rr{DR}\right)+
\frac{1}{1+R_\rr{dark}}\frac{1}{\Gamma_\rr{DR-DM}}\times \\
&\times \left[- \frac{\ddot{a}}{a} \theta_\rr{DM}
- \frac{\dot{a}}{a}\left(k^2\frac{1}{2}\delta_\rr{DR}+k^2\psi\right)+k^2\left(c_\rr{DM}^2 \dot{\delta}_\rr{DM}-\frac{1}{4}\dot{\delta}_\rr{DR}\right)\right],
\end{align*}
where $\dot{}$ denotes the derivative with respect to conformal time.
The slip is then plugged into the exact equations for the DM and DR dipole moments $\dot{\theta}_{\rr DM}$ and $\dot{\theta}_{\rr DR}$.

%% file: Results.tex
\section{Results}
\label{sec:results}
With the method implemented above, we have used {\sc MontePython}~\cite{Audren:2012vy,Brinckmann:2018cvx}, interfaced with our modified {\sc class} version, in its default Metropolis Hastings mode, to perform parameter scans on the combination of $\{\omega_\rr{b},\ \omega_\rr{cdm},\ \log(10^{10}A_\rr{s}),\ n_\rr{s},\ \tau_\rr{reio },\ H_0,\ \xi,\ a_\rr{dark }\}$, for the ETHOS models with $n=4$, $n=2$, and $n=0$ (corresponding to different powers of the temperature dependence of the co-moving interaction rate $\Gamma\propto T^n$). For $n=4$ and $n=2$ we assume DR to be free-streaming and we neglect the impact of DR self-interactions~\cite{Archidiacono:2017slj} (i.e. $\Gamma_\mathrm{DR-DR}=0$, {\tt b\_dark}$=0$), while for $n=0$ we assume DR to behave like a fluid (i.e. $\Gamma_\mathrm{DR-DR}\rightarrow \infty$, {\tt idr\_nature}={\tt fluid}).
For the final case, we have also investigated the impact of changing our choice of parameters to match the NADM model discussed in~\cite{Lesgourgues:2015wza,Buen-Abad:2017gxg}, thus giving us  $\{\omega_\rr{b},\ \omega_\rr{cdm},\ \log(10^{10}A_\rr{s}),\ n_\rr{s},\ \tau_\rr{reio },\ H_0,\ \Delta N_\rr{fluid},\ \Gamma_0\}$.

\begin{figure}[t]
\begin{center}
\begin{tabular}{cc}
\includegraphics[width=0.5\linewidth]{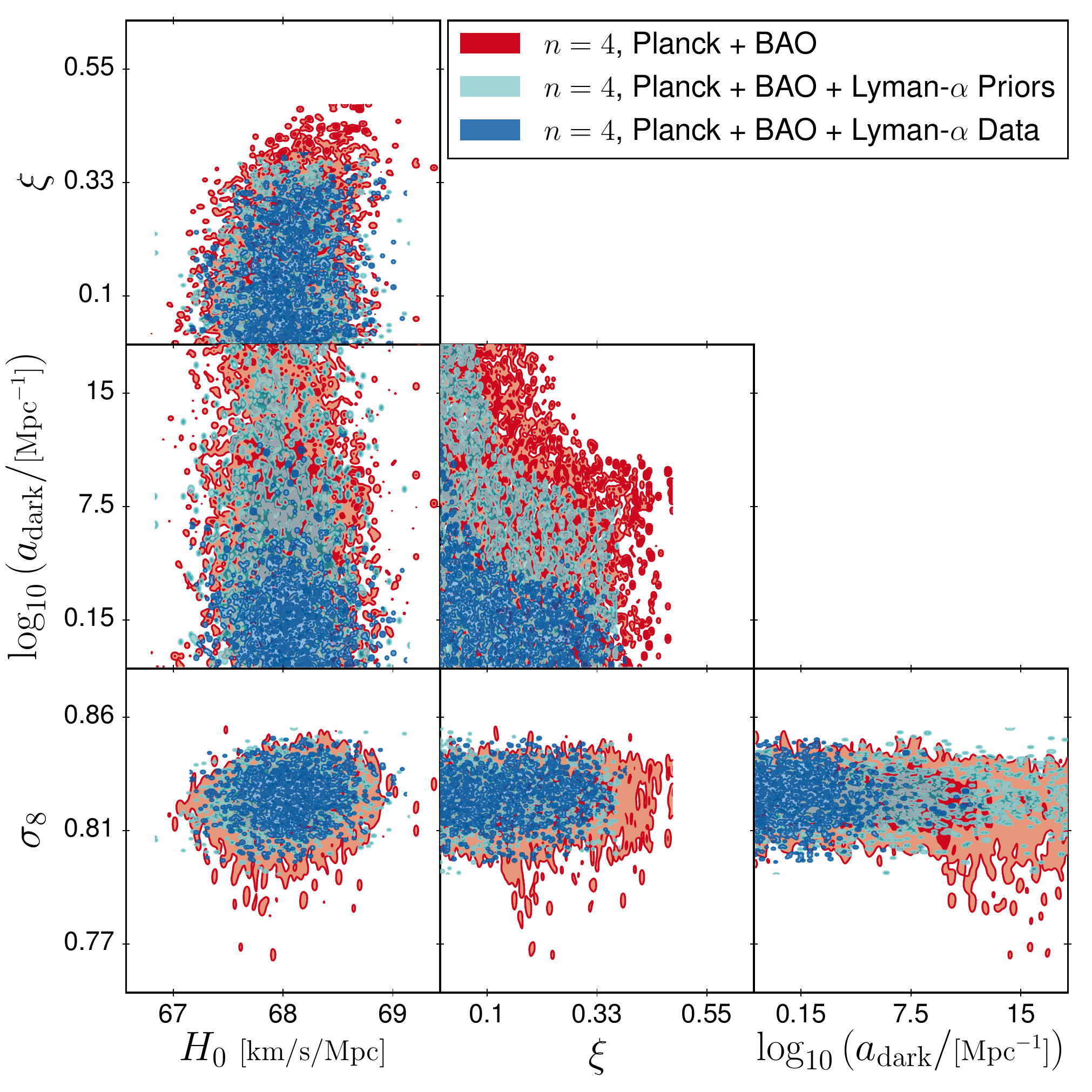}&
\includegraphics[width=0.35\linewidth]{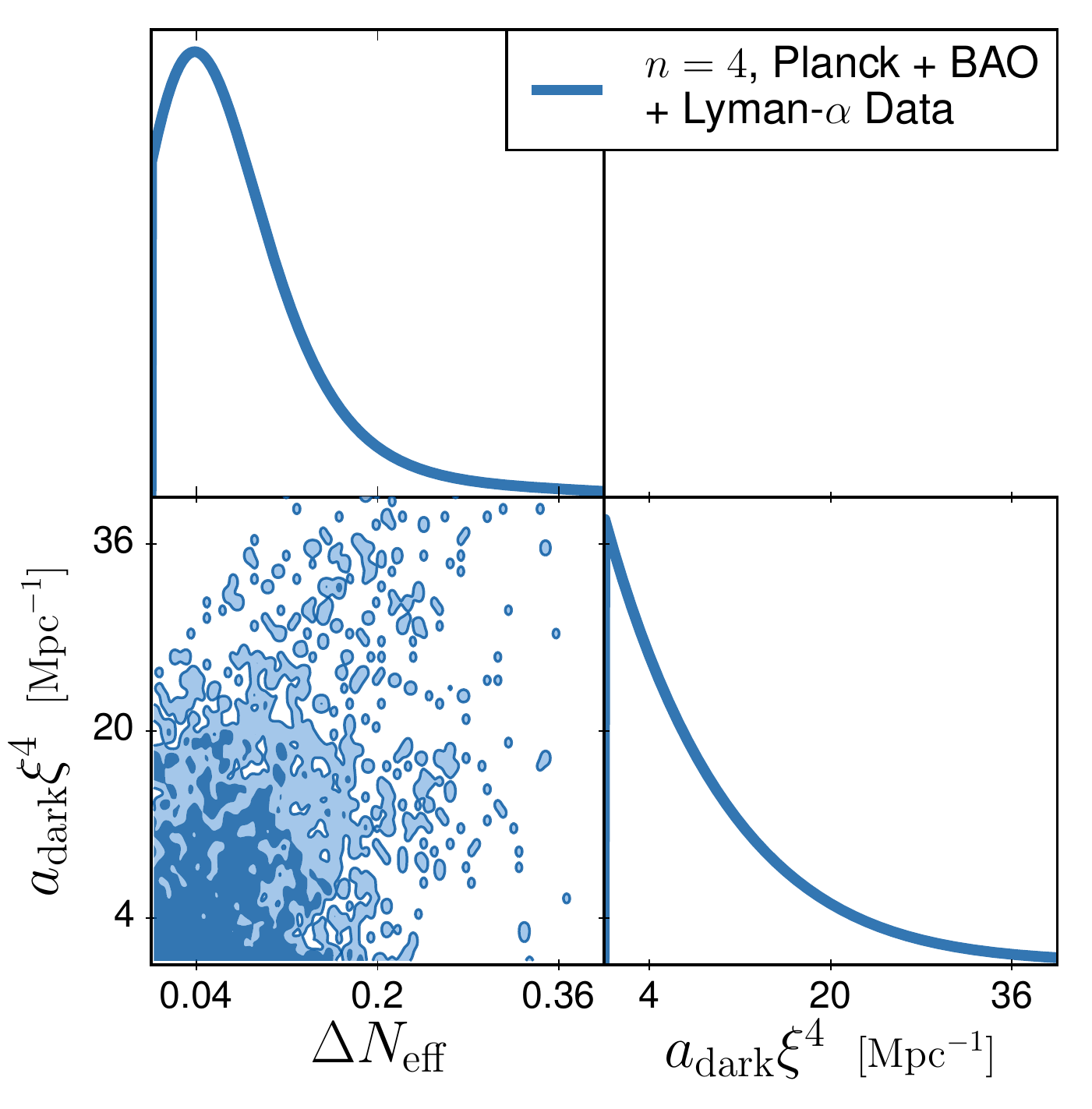}
\end{tabular}
\caption{\emph{(Left)} Two-dimensional posterior distributions for all main parameters for the $n = 4$ case, with Planck + BAO (red), Planck + BAO + Lyman-$\alpha$ Data (dark blue), and the Lyman-$\alpha$ Prior check run explained in the text (light blue), when running with a flat prior on $\xi$ and logarithmic prior on $a_\mathrm{dark}$. The smoothing has deliberately been turned off to show the sharp boundaries of the preferred regions more clearly. \emph{(Right)} Posterior distributions when using linear priors on $\Delta N_\mathrm{eff}$ and $a_\mathrm{dark}\xi^4$.}
\label{fig:n4}
\end{center}
\end{figure} 

For each of the studied ETHOS models, we performed MCMC analysis for two different data combinations:
\begin{itemize}
\item \textbf{Planck + BAO}: This is the combination of Planck 2015 high-$\ell$ TT+TE+EE, low-$\ell$ data~\cite{Aghanim:2015xee} and Planck 2015 lensing data~\cite{Ade:2015zua}. We further add BAO data, using measurements of $D_V/r_{\rm drag}$ by 6dFGS at $z = 0.106$~\cite{Beutler:2011hx} by SDSS from the MGS galaxy sample at $z = 0.15$~\cite{Ross:2014qpa}, and additionally by BOSS from the CMASS and LOWZ galaxy samples of SDSS-III DR12 at $z = 0.2 - 0.75$~\cite{Alam:2016hwk}.
\item \textbf{Planck + BAO +  Lyman-$\alpha$}: Same as above, with the additional Lyman-$\alpha$ likelihood described in section~\ref{sec:methods}.
\end{itemize}
The results for the different cases are discussed below.

\subsection{ETHOS $n=4$ model}
The underlying particle physics model that leads to the $n=4$ temperature dependence of the comoving interaction rate is represented by fermionic relativistic particles (DR), e.g. sterile neutrinos, interacting with DM particles through a new massive boson mediator of a new $U(1)$ broken symmetry.
Given the negligible impact of DR self-interactions induced by these processes on the matter power spectrum~\cite{Archidiacono:2017slj}, we set $\Gamma_\mathrm{DR-DR}$ to $0$. 
The results of our MCMC runs for this model are shown in Fig.~\ref{fig:n4} and Table~\ref{table:n4}, for both of the dataset combinations mentioned above.

\begin{table}[t]
\begin{center}
\begingroup
\renewcommand{\arraystretch}{1.5} 
\begin{tabular}{|c|c|c|c|}
\hline
& $\Lambda$CDM & \multicolumn{2}{c|}{ETHOS $n=4$}\\ 
parameter & Planck+BAO & Planck + BAO & + Lyman-$\alpha$\\
\hline \hline
$100~\omega_\rr{b }$ &  $2.219_{-0.014}^{+0.013}$ & $2.221_{-0.015}^{+0.015}$  & $2.222_{-0.014}^{+0.017}$ \\ 
$\omega_\rr{cdm }$ &  $0.1192_{-0.0010}^{+0.0011}$ & $0.1195_{-0.0014}^{+0.0011}$ &$0.1192_{-0.0010}^{+0.0011}$ \\ 
$\log(10^{10}A_\rr{s })$ & $3.050_{-0.023}^{+0.023}$ &  $3.053_{-0.023}^{+0.025}$&$3.057_{-0.024}^{+0.024}$ \\ 
$n_\rr{s }$ & $0.9618_{-0.0041}^{+0.0042}$ & $0.9622_{-0.0045}^{+0.0044}$ & $0.9626_{-0.0037}^{+0.0044}$ \\ 
$\tau_\rr{reio }$ & $0.060_{-0.012}^{+0.012}$ & $0.061_{-0.013}^{+0.013}$ & $0.063_{-0.013}^{+0.013}$\\ 
$H_0 \ / \left[{\rm km/(s \,Mpc)}\right]$ & $67.94_{-0.49}^{+0.46}$ & $68.06_{-0.54}^{+0.52}$& $68.09_{-0.48}^{+0.46}$\\ 
$\sigma_8$ &  $0.8234_{-0.0090}^{+0.0085}$ & $0.823_{-0.013}^{+0.024}$ & $0.826_{-0.009}^{+0.010}$\\
$n_\rr{eff}$ &  $-2.3080_{-0.0035}^{+0.0034}$ & $-2.9_{-22.1}^{+4.3}$ &  $-2.3070_{-0.0035}^{+0.0039}$\\
$\xi$ & -- & $<0.40$  & $<0.38$ \\ 
$\log_{10}(a_\rr{dark } \ / \left[{\rm Mpc}^{-1}\right])$ & -- & n.l. &  $<6.8$\\ 
$\Delta\chi^2$ & -- & $0$ & $-3.62$\\
\hline
\hline
$\Delta N_\mathrm{eff}$ & -- & -- & $<0.23$ \\
$a_\mathrm{dark}\xi^4 /   \left[{\rm Mpc}^{-1}\right]$ & -- & -- & $<30$ \\
\hline
\end{tabular}
\endgroup
\end{center}
\caption{Preferred regions at the 68\,\% Confidence Level (C.L.) (or at the 95\,\% C.L. in the case of upper bounds) for the parameters of the ETHOS $ n = 4 $ case, both with Planck + BAO and Planck + BAO + Lyman-$\alpha$. With the first dataset, the interaction parameter is not bounded within the prior range. The $\Delta\chi^2$ is given with respect to $\Lambda$CDM with the same datasets. The last two rows show the results obtained with linear priors on $\Delta N_\mathrm{eff}$ and $a_\rr{dark}\xi^4$ using the second dataset. Entries with ``n.l.'' means that there is no upper limit within the prior range, while -- means that the parameter is not present.}
\label{table:n4}
\end{table}

\vspace{0.5cm}

\noindent {\it CMB constraints.}~We expect a clear degeneracy between the amount of DR $\xi$ and the interaction strength $a_\mathrm{dark }$, because the data should remain compatible with DM interacting either strongly with a small amount of DR or barely with a large amount of DR. To capture this behaviour, we chose to use a flat prior on $\log_{10}(a_\mathrm{dark })$ in the range $[-3,20]$. Indeed, a linear prior on $a_\mathrm{dark }$ would only give weight to the region with a high interaction rate and thus a tiny DR density. This would lead to very strong bounds on $\xi$ that would not reflect the fact that the data is perfectly compatible with values up to $\xi \sim 0.40$. 

In the middle plot of the left panel of  Fig.~\ref{fig:n4}, we can see the expected degeneracy between $\xi$ and $\log_{10}(a_\mathrm{dark })$. The results of MCMC runs are usually plotted as smoothed contour plots. In this particular work, we choose instead to plot the non-smoothed density of points in the chains\footnote{In practice this is achieved by analysing the chains with a high number of bins (one hundred).}, in order to precisely visualise the edges of the region preferred by the data. The Planck + BAO allowed region has two sharp edges set by the data rather than the priors:
\begin{itemize}
\item a vertical line corresponding to the maximum allowed value of $\xi$ (and therefore $\Delta N_\mathrm{eff}$) in the ETHOS $n=4$ model. We find $\xi< 0.40$ (95~\%~C.L.), which is consistent within $1\sigma$ with the bound obtained in Ref.~\cite{Archidiacono:2017slj}, with our bounds being slightly tighter. This small difference can be attributed to our inclusion of the lensing and BAO likelihoods, which were not included in the previous study. This can be translated into $\Delta N_\mathrm{eff}<0.10$, but the latter result must be taken with a grain of salt because it derives from a flat prior on $\xi$. Later in this section we will report another bound obtained with a flat prior on $\Delta N_\mathrm{eff}$. The physical interpretation of this boundary is that the CMB data is incompatible with too much DR, even when the latter is self-interacting. This is caused by various effects, the dominant one being the influence of the amount of extra radiation on the CMB damping tail~\cite{Hou:2011ec}. DR has other effects on the scale and amplitude of the acoustic peaks that depend on the rate of DR self-interactions and DR-DM interaction~\cite{Archidiacono:2017slj}: thus the bound found in this case is specific to the ETHOS $n=4$ model, and in principle different from what one would obtain in a plain $\Lambda$CDM$+N_\mathrm{eff}$ fit with only free-streaming relativistic relics.
\item a roughly hyperbolic boundary, corresponding physically to the limit set by the CMB on the effect of the DM-DR interaction. In particular, a too large rate $\Gamma_\mathrm{DM-DR}$ implies that DM develops a fast mode~\cite{Weinberg:2002kg,Voruz:2013vqa} that influences the CMB power spectrum, with a suppression of the clustering of the baryon-photon fluid~\cite{Archidiacono:2017slj,Cyr-Racine:2015ihg,Cyr-Racine:2013fsa}.
\end{itemize}
We obtain no upper bound on $\log_{10}(a_\mathrm{dark })$, since in the limit of small DR density the DM-DR and DR-DM interaction rates can be arbitrarily high. Thus the allowed region extends up to our upper prior boundary $\log_{10}(a_\mathrm{dark })\leq20$.

For the other cosmological parameters, error bars are slightly larger than for the $\Lambda$CDM model with the same data combination, but smaller than for the $\Lambda$CDM$+N_\mathrm{eff}$ model. This arises from several reasons: our flat prior on $\xi$ gives more weight to small values of $\Delta N_\mathrm{eff}$; we only allow $N_\mathrm{eff}$ to increase beyond 3.046, while a run with a flat prior on $N_\mathrm{eff}$ would return $N_\mathrm{eff} = 2.98 \pm 0.18$ (68~\%~C.L.) \cite{Ade:2015xua,PLA2015};  and in our model, increasing $\Delta N_\mathrm{eff}$ comes at the price of introducing DM-DR interaction effects not favoured by the data. In any case we see that the ETHOS $n=4$ model offers no clear opportunities to accommodate the high value of $H_0$ \cite{Riess:2019cxk} and/or the low value of $\sigma_8$ hinted by some datasets \cite{Hildebrandt:2018yau,Abbott:2017wau,Joudaki:2019pmv}.

\vspace{0.5cm}

\noindent {\it Lyman-$\alpha$ constraints.}~With the addition of the Lyman-$\alpha$ likelihood, we obtain approximately the same bound on $\xi< 0.38$ (95\,\% C.L.), as the number of additional relativistic degrees of freedom is already well-constrained by CMB data. Instead the upper limit on the interaction rate shrinks by about ten orders of magnitude, because DM-DR interactions result in a suppression of the small-scale matter power spectrum strongly constrained by Lyman-$\alpha$ data. Quantifying this effect is the main goal of this paper. Figure~\ref{fig:Tk_lin} already showed that a larger value of $a_\mathrm{dark}$ could potentially be compensated by a smaller value of $\xi$ leading to the same cut-off scale. Indeed, we checked explicitly that the edge of the allowed region is a curve of constant $a_\mathrm{dark} \xi^4$. This behaviour was expected because the term that accounts for interactions in the DM Euler equation has a coefficient $\Gamma_\mathrm{DM-DR} \propto \rho_\mathrm{DR} \, \Gamma_\mathrm{DR-DM} \propto  a_\mathrm{dark} \xi^4$.

This run gives an upper bound $\log_{10}(a_\mathrm{dark }/{\rm Mpc}^{-1})< 6.8$ (95~\%~CL) that is strongly prior dependent. Indeed, since $a_\mathrm{dark }$ is compatible with zero, upper bounds on $\log_{10}(a_\mathrm{dark})$ are inevitably influenced by the choice of a lower prior boundary on this parameter. Moreover, the data are compatible with arbitrarily large values of $a_\mathrm{dark }$ for arbitrarily small $\xi$'s, such that the bound would entirely disappear if we had chosen a logarithmic prior on $\xi$.

The analysis with flat priors on $\xi$ and $\log_{10}(a_\mathrm{dark })$ is particularly useful for identifying the physical mechanisms responsible for the various bounds. It allowed us to check that the data are mostly sensitive to the effect of the density of extra radiation, proportional to $\Delta N_\mathrm{eff}$, and of the DM-DR rate $\Gamma_\mathrm{DM-DR}$, parametrised by $a_\mathrm{dark} \xi^4$. Therefore, the most informative and robust way to formulate our final results is to quote bounds on ($\Delta N_\mathrm{eff}$, $a_\mathrm{dark} \xi^4$) assuming flat priors on these parameters. 

We thus performed another MCMC run with such a choice of priors. The results are shown in the right panel of Fig.~\ref{fig:n4}. Our final results for the $n=4$ ETHOS model are summarised by the 95~\%~upper bounds $\Delta N_\mathrm{eff}<0.23$ and $a_\mathrm{dark} \xi^4<30 \, {\rm Mpc}^{-1}$. 
The upper limit of the Bayesian confidence interval for $\Delta N_\mathrm{eff}$ is slightly stronger than for a $\Lambda$CDM$+N_\mathrm{eff}$ model with extra free-streaming relativistic relics and Planck+BAO data, $\Delta N_\mathrm{eff}<0.28$ (95~\%~C.L., see \cite{Ade:2015xua,PLA2015}), because in our case models with $\Delta N_\mathrm{eff}>0$ also come with DM-DR interaction effects that are not favoured by the data.
Knowing the upper bound on $a_\mathrm{dark} \xi^4$ is convenient for model building. A typical particle-physics-motivated model would predict a given value of $\xi$ (related to the physics of the dark sector and to its interactions with the visible sector). In such a case one can immediately conclude that the Lyman-$\alpha$ data impose a maximum value on the scattering rate $a_\mathrm{dark}$ given by $30\, \xi^{-4} \, {\rm Mpc}^{-1}$. 

It is important to check that our results are actually driven by the Lyman-$\alpha$ data, and not by the restrictions imposed on the small-scale matter power spectrum by the method implemented in our likelihood, that we described in section~\ref{sec:methods}. For this purpose, we also performed a run with the Planck + BAO likelihoods combined with a modified version of the Lyman-$\alpha$ likelihood that returns a constant value if the power spectrum passes all of the sanity checks, and a zero likelihood otherwise. Thus this run relies on the Planck + BAO data and on the Lyman-$\alpha$ likelihood prior, but not on the  Lyman-$\alpha$ data. It allows us to derive regions of validity for our implementation. We call it ``Planck + BAO + Lyman-$\alpha$ Prior'' and its results are also shown in the left panel of Fig.~\ref{fig:n4}. If the edge of the allowed region was similar in the Lyman-$\alpha$ Prior and Lyman-$\alpha$ Data runs, we would know that our bounds are driven by the applicability of the method and not by the data. This is not the case, as we can clearly see when comparing the dark and light blue regions in  Fig.~\ref{fig:n4}. As such, we conclude that the sanity checks of our implementation impose no further restriction besides the region that is already excluded by other means.
 
Furthermore, when adding the Lyman-$\alpha$ likelihood, our error bars on $n_\mathrm{eff}$, which is the slope of the Lyman-$\alpha$ spectrum, are greatly reduced. This comes mainly from our improved bound on $a_\mathrm{dark}$; when the interaction strength is allowed to vary over many orders of magnitude, our $P(k)$ is not monotonic, and thus $n_\mathrm{eff}$ can assume any value, both negative and positive (if the corresponding $k$ value is, for example, just after the first oscillation in $P(k)$).

The inclusion of Lyman-$\alpha$ data tightens the error bars on $\sigma_8$, while the mean value is not significantly affected. The mean value and error bars of $H_0$ are not impacted by the addition of  Lyman-$\alpha$ data for this model. The bounds for both parameters are in very close agreement with those obtained for a standard $\Lambda$CDM model with the same datasets. 

Finally, the $\chi^2$ obtained in the Planck + BAO case is not any better than for the vanilla $\Lambda$CDM model, while the addition of Lyman-$\alpha$ data brings it down by $\Delta\chi^2=-3.6$. Considering that the model features two additional parameters, we conclude that interacting DM-DR models provide a fit of Planck + BAO + Lyman-$\alpha$ as good as $\Lambda$CDM.

\subsection{ETHOS $n=2$ model}

\begin{figure}[t]
\begin{center}
\begin{tabular}{cc}
\includegraphics[width=0.5\linewidth]{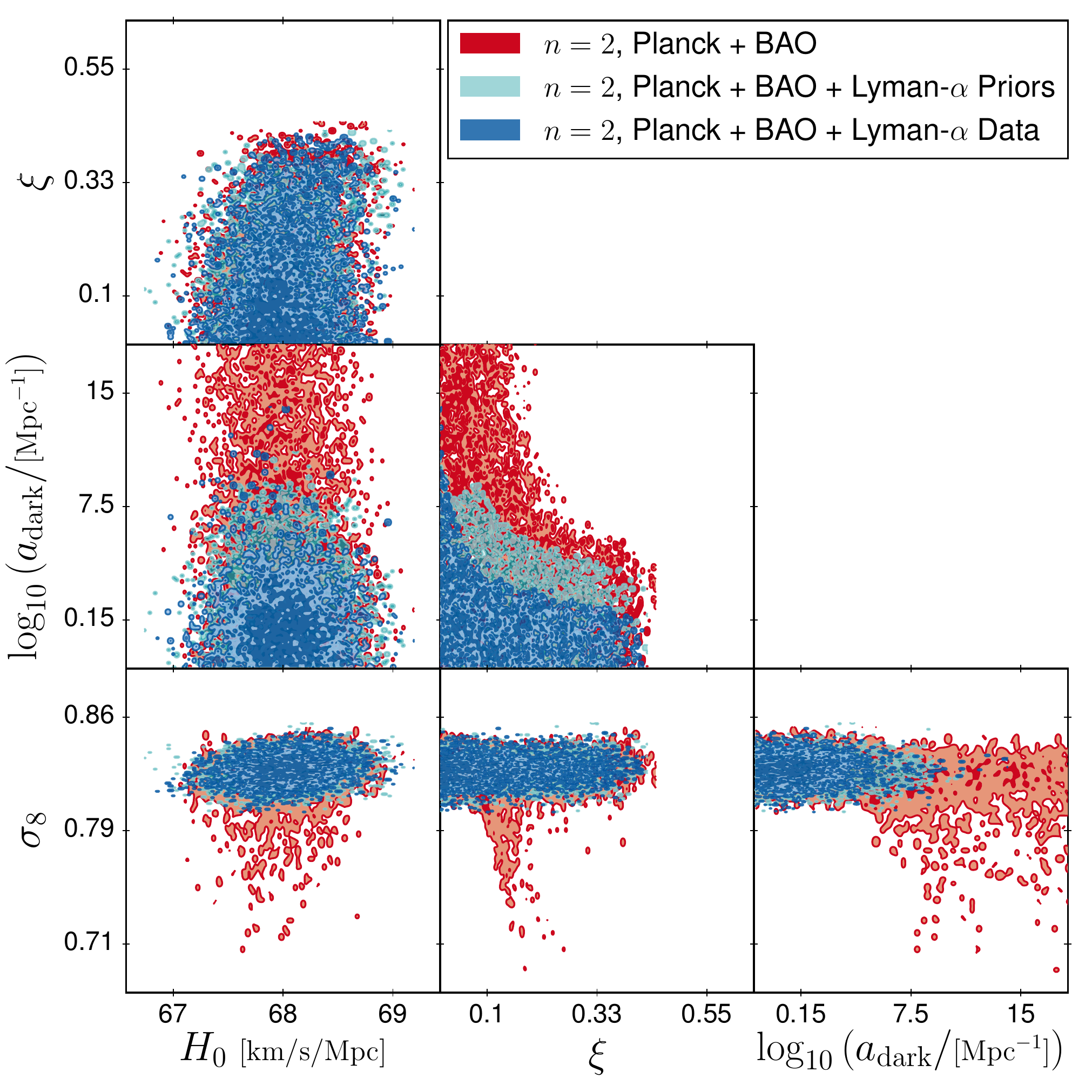}&
\includegraphics[width=0.35\linewidth]{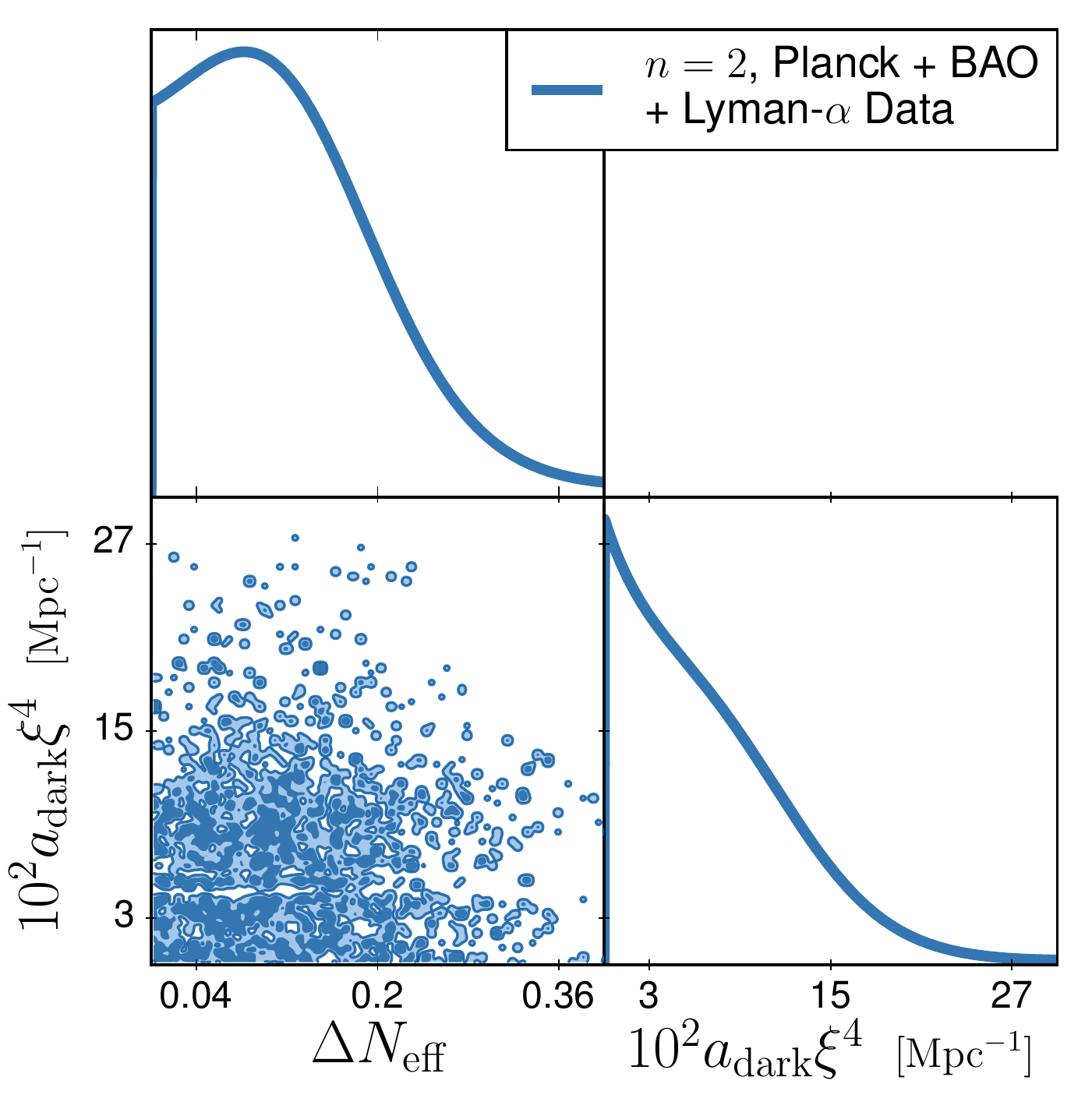}
\end{tabular}
\caption{\emph{(Left)} Two-dimensional posterior distributions for all main parameters for the $n = 2$ case, with Planck + BAO (red), Planck + BAO + Lyman-$\alpha$ Data (dark blue), and the Lyman-$\alpha$ Prior check run explained in the text (light blue), when running with a flat prior on $\xi$ and  logarithmic prior on $a_\mathrm{dark}$. The smoothing has deliberately been turned off to show the sharp boundaries of the preferred regions more clearly. \emph{(Right)} Posterior distributions when using linear priors on $\Delta N_\mathrm{eff}$ and $10^2 a_\mathrm{dark}\xi^4$.
}
\label{fig:n2}
\end{center}
\end{figure}

The scenario where the comoving scattering rate of DR off DM scales like $T^2$ can be realised e.g. with 4-point contact-only interaction. As for $n=4$, we neglect the subdominant contribution of DR self-interactions.

\vspace{0.5cm}

\noindent {\it CMB constraints.}~The results of our MCMC run with Planck+BAO data and for the $n=2$ case are shown in Fig.~\ref{fig:n2} and Table~\ref{table:n2}. Once more, the middle plot in the left panel of Fig.~\ref{fig:n2} shows that the data impose two limitations on the ETHOS parameter: an upper bound $\xi<0.43$ at the 95~\%~C.L. and a hyperbolic-shaped limit on ($\xi$, $a_\mathrm{dark}$). 

For other parameters, the preferred intervals only widen moderately with respect to the $\Lambda$CDM model, excepted for $\sigma_8$ which is compatible with much smaller values. 
The contour plot for ($\xi$, $\sigma_8$) shows a degeneracy direction allowing to reach such small values for specific values of $\xi$ and a large interaction rate $a_\mathrm{dark} > 10^7$. The degeneracy is captured by the relation $\sigma_8 \simeq 0.823-210 \, \xi^4$, and stretches down to $\sigma_8 = 0.75$ for $\xi \simeq 0.14$. It is potentially interesting to explain the low value of $\sigma_8$ returned by several data on cosmic shear and cluster counts, but we will not investigate it in details because this region will be excluded in the next paragraph by Lyman-$\alpha$ bounds on the interaction rate. Once this region is ignored, we find that the ETHOS $n=2$ model does not offer opportunities to accommodate larger $H_0$ or smaller $\sigma_8$ values than $\Lambda$CDM.

\vspace{0.5cm}

\noindent {\it Lyman-$\alpha$ constraints.}~Like for $n=4$, the inclusion of Lyman-$\alpha$ data marginally affects the bound on $\xi$, but considerably strengthens the upper limit on the interaction rate, which is given once more by a line of constant $a_\mathrm{dark} \xi^4$. This limit is stronger than in the $n=4$ case by about two orders of magnitude. We checked explicitly that the suppression in the matter power spectrum takes place roughly at the same scale when we change $n$ and keep the same $10^{-n} a_\rr{dark} \xi^4$. This is consistent with the fact that the scales constrained by our Lyman-$\alpha$ data crossed the Hubble scale roughly around $z \simeq 10^6$, and have been suppressed according to the rate $\Gamma_\mathrm{DM-DR}(z)$ evaluated at that time. Equations (\ref{eq:DRDM},\ref{eq:DMDR}) show that up to constant numbers, 
\begin{equation}
\Gamma_\mathrm{DM-DR}(z) 
\propto (1+z) \left( \frac{1+z}{1+z_d} \right)^n a_\mathrm{dark} \xi^4
\label{eq:gammadmdrz}
\end{equation}
with $z_d=10^7$, implying 
\begin{equation}
\Gamma_\mathrm{DM-DR}(10^6) \propto 10^{6-n}  a_\mathrm{dark} \xi^4~.
\end{equation}
Thus it is normal that the Lyman-$\alpha$ dataset provides comparable limits on the combination $(10^{-n}  a_\mathrm{dark} \xi^4)$ for all $n$'s, and that limits on $a_\mathrm{dark} \xi^4$ become one hundred times stronger when $n$ decreases by two.

We find a bound $\xi < 40$ (95~\%~C.L.) very similar to that in the $n=4$ case, while the bound $\log_{10} (a_\mathrm{dark} / [\mathrm{Mpc}^{-1}]) < 8.4$  (95~\%~C.L.) should again be taken with great care due to its strong dependence on the choice of a linear prior for $\xi$ and on the lower prior edge for $\log_{10} (a_\mathrm{dark})$. We thus switch to linear priors on the parameters directly related to the physical effects probed by the data, and obtain our final results for the ETHOS $n=2$ model: $\Delta N_\mathrm{eff}<0.29$ and $10^2 a_\mathrm{dark}\xi^4<18 \, \mathrm{Mpc}^{-1}$ (95~\%~C.L.). The first bound is identical to what is obtained when fitting Planck+BAO with a $\Lambda$CDM$+N_\mathrm{eff}$ model.

Once again we performed a ``Planck + BAO + Lyman-$\alpha$ Prior'' run to check that our bounds do not come from the limitations of the method. In this case, if we compare the $\xi- \log(a_\mathrm{dark })$ posteriors for the Lyman-$\alpha$ Prior and Lyman-$\alpha$ Data runs in the left panel of Fig.~\ref{fig:n2}, we see that for $\xi>0.05$ our constraints are really derived from the data rather than from the range of validity of our method. This is not true any more in a very small region with $\xi<0.05$, where the two contours overlap. This is because for these models, the $\{\alpha, \beta, \gamma \}$-parametrisation is not accurate.
However $\xi<0.05$ implies a tiny DR density $\Delta N_\mathrm{eff}<2 \cdot 10^{-5}$. This small region is not very interesting for model building, because such tiny values are difficult to motivate theoretically (for instance, they may derive from a DR particle decoupling from thermal equilibrium with standard model  particles when the number of relativistic degrees of freedom is unusually large, $g_* \sim {\cal O} (10^4)$). Also, even if our method was improved in order to deal correctly with this corner of the parameter space, there would be no reason for the 95~\%~C.L. upper bound on ($\xi$, $a_\mathrm{dark}$) to be different from $10^2 a_\mathrm{dark}\xi^4 = 18$, since the shape of this limit can be inferred from simple analytic arguments. Thus we can safely extrapolate it below $\xi = 0.05$. Finally, we should note that this minor issue is irrelevant when running with a flat prior on $\Delta N_\mathrm{eff}$, since with such a prior it affects a completely negligible fraction of the preferred region volume.


\begin{table}[t]
\begin{center}
\begingroup
\renewcommand{\arraystretch}{1.5} 
\begin{tabular}{|c|c|c|c|}
\hline
& $\Lambda$CDM & \multicolumn{2}{c|}{ETHOS $n=2$}\\ 
parameter & Planck+BAO & Planck + BAO & + Lyman-$\alpha$\\
\hline \hline 
$100~\omega_\rr{b }$ & $2.219_{-0.014}^{+0.013}$ & $2.220_{-0.014}^{+0.014}$   & $2.220_{-0.016}^{+0.014}$ \\ 
$\omega_\rr{cdm }$ & $0.1192_{-0.0010}^{+0.0011}$ & $0.1195_{-0.0013}^{+0.0011}$ & $0.1194_{-0.0011}^{+0.0011}$ \\ 
$\log(10^{10}A_\rr{s })$ & $3.050_{-0.023}^{+0.023}$ & $3.053_{-0.025}^{+0.025}$  & $3.051_{-0.024}^{+0.023}$\\ 
$n_\rr{s }$ & $0.9618_{-0.0041}^{+0.0042}$ & $0.9621_{-0.0043}^{+0.0044}$ & $0.9618_{-0.0043}^{+0.0039}$\\ 
$\tau_\rr{reio }$ & $0.060_{-0.012}^{+0.012}$ & $0.061_{-0.013}^{+0.013}$  & $0.059_{-0.013}^{+0.013}$ \\ 
$H_0 \ / \left[{\rm km/(s \,Mpc)}\right]$ & $67.94_{-0.49}^{+0.46}$ & $68.02_{-0.51}^{+0.51}$  &$67.99_{-0.51}^{+0.51}$\\ 
$\sigma_8$ & $0.8234_{-0.0090}^{+0.0085}$ & $0.819_{-0.017}^{+0.021}$ & $0.8244_{-0.0095}^{+0.0088}$\\
$n_\rr{eff}$ & $-2.3080_{-0.0035}^{+0.0034}$ & $-2.9_{-3.5}^{+7.0}$  &$-2.3080_{-0.0037}^{+0.0034}$ \\
$\xi$ & -- & $<0.43$ & $ < 0.40$ \\ 
$\log_{10}(a_\rr{dark } \ / \left[{\rm Mpc}^{-1}\right])$ & -- & n.l. & $ < 8.4$\\ 
$\Delta\chi^2$ & -- & $0$ & $-0.12$\\
\hline
\hline
$\Delta N_\mathrm{eff}$ & -- & -- &$<0.29$ \\
$10^{2} a_\mathrm{dark}\xi^4/   \left[{\rm Mpc}^{-1}\right]$ & -- & -- &$<18$ \\
\hline
\end{tabular}
\endgroup
\end{center}
\caption{Preferred ranges at the 68\,\% Confidence Level (or 95\,\% upper bound in some cases) for all relevant parameters for the ETHOS $ n = 2 $ case, both with Planck + BAO and Planck + BAO + Lyman-$\alpha$. With the first dataset, the interaction parameter is not bounded within the prior range. The $\Delta\chi^2$ is given with respect to $\Lambda$CDM with the same datasets. The last two rows show the results obtained with linear priors on $\Delta N_\mathrm{eff}$ and $10^2a_\mathrm{dark}\xi^4$ using the second dataset.}
\label{table:n2}
\end{table}

Like for the $n=4$ case, we obtain a significantly tighter bound on $n_\mathrm{eff}$, while the mean value and error bars of $H_0$ are not impacted by the addition of Lyman-$\alpha$ data. The preferred intervals for $H_0$ and $\sigma_8$ are very close to those of the $\Lambda$CDM model. For both data combinations, the difference obtained in the $\Delta\chi^2$ with respect to the base $\Lambda$CDM are negligible, thus we once again find no preference for the interacting DM-DR models.

\subsection{ETHOS $n=0$ model}
\begin{figure}[t]
\begin{center}
\begin{tabular}{cc}
\includegraphics[width=0.5\linewidth]{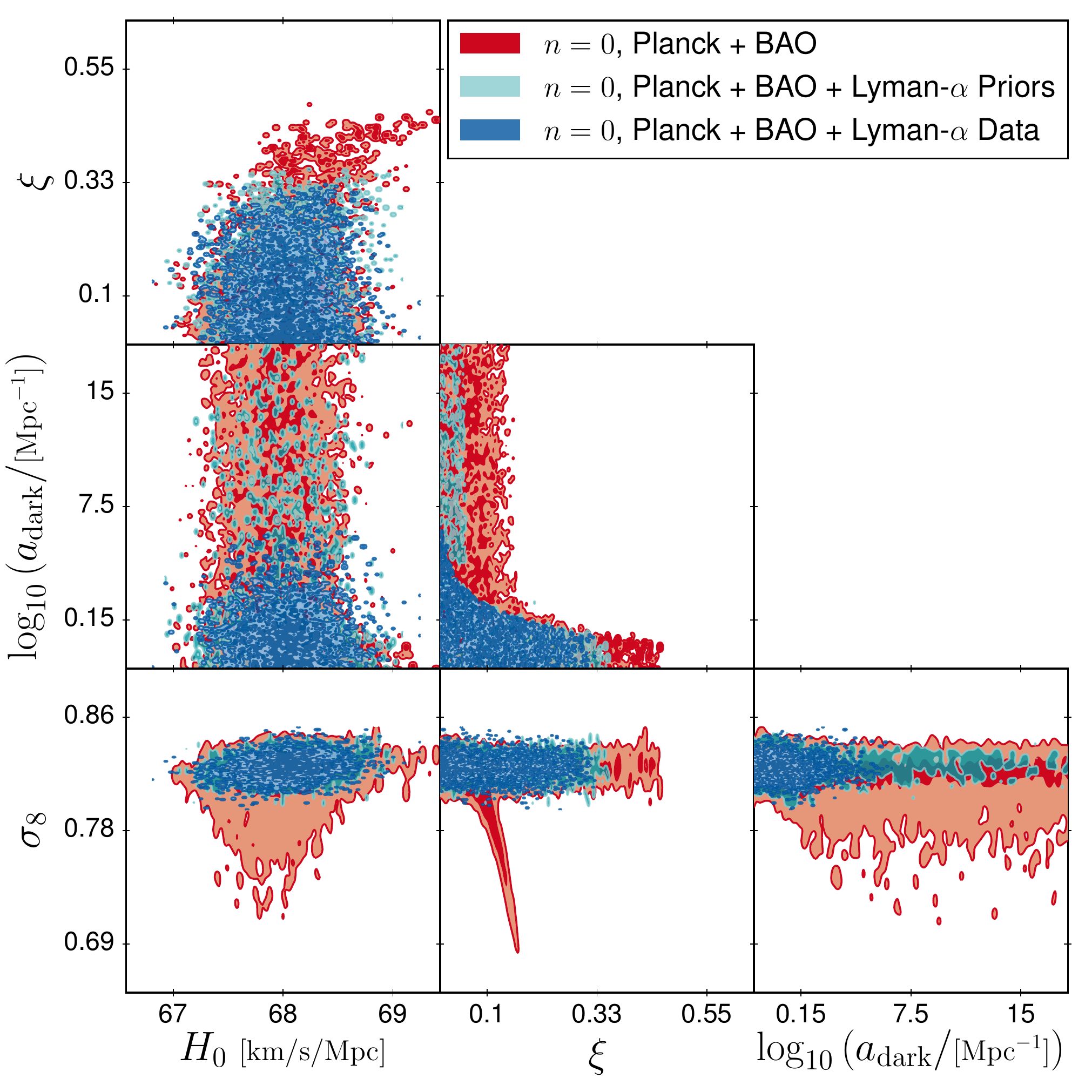}&
\includegraphics[width=0.35\linewidth]{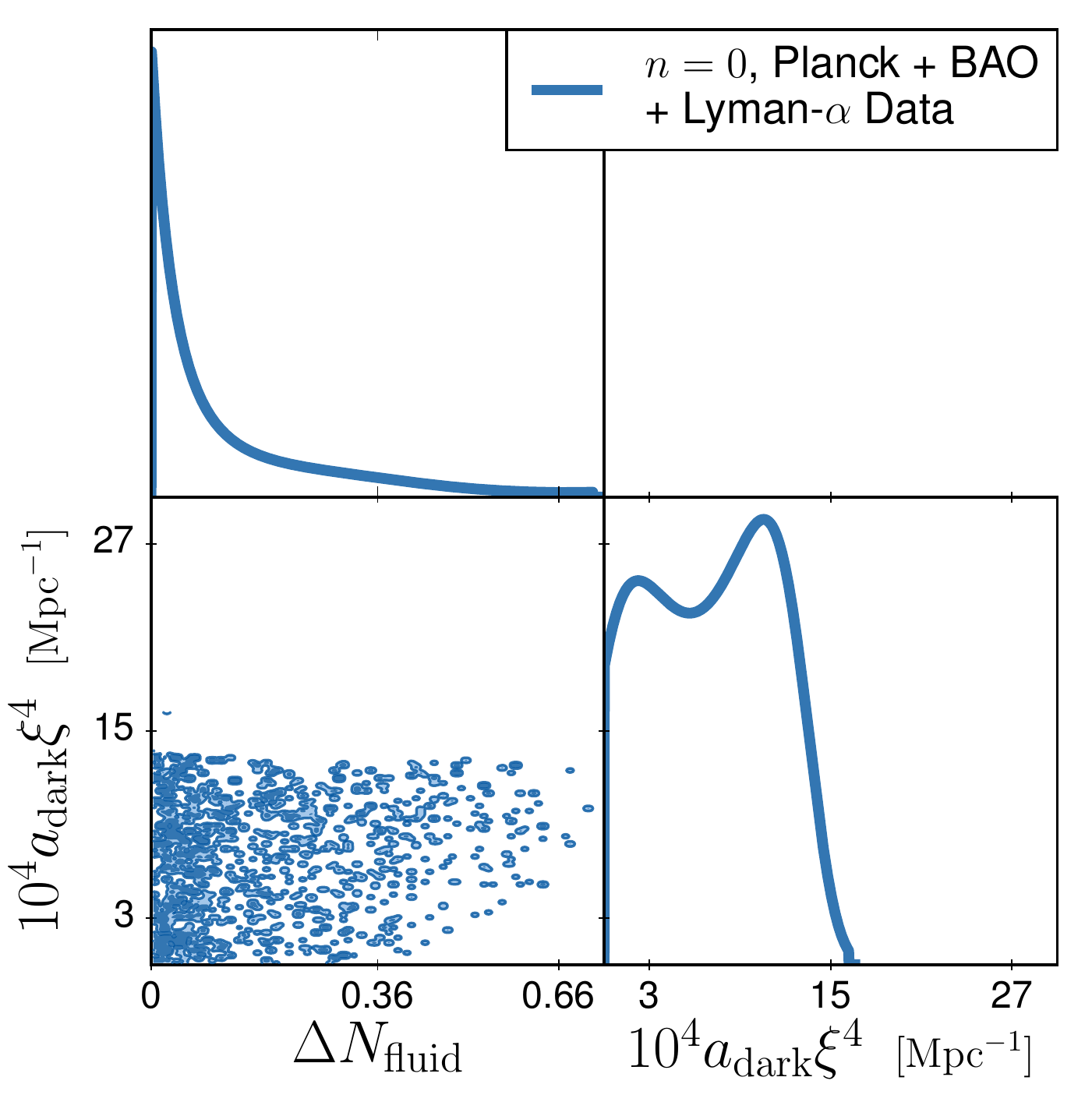}
\end{tabular}
\caption{\emph{(Left)} Two-dimensional posterior distributions for all main parameters for the $n = 0$ case, with Planck + BAO (red), Planck + BAO + Lyman-$\alpha$ Data (dark blue), and the Lyman-$\alpha$ Prior check run explained in the text (light blue), when running with a flat prior on $\xi$ and  logarithmic prior on $a_\mathrm{dark}$. The smoothing has deliberately been turned off to show the sharp boundaries of the preferred regions more clearly. \emph{(Right)} Posterior distributions when using linear priors on $\Delta N_\mathrm{fluid}$ and $10^4 a_\mathrm{dark}\xi^4$.
}
\label{fig:n0}
\end{center}
\end{figure}

The $n=0$ case is motivated by several particle physics setups in which the DM-DR momentum transfer rate with respect to proper time scales like $T^2$, meaning that the ETHOS rate $\Gamma_\mathrm{DR-DM}$ is constant. This occurs for instance in the Non-Abelian Dark Matter (NADM) scenario, in which DM particles are charged under a dark non-abelian symmetry whose dark gluons play the role of DR (see e.g.~\cite{Buen-Abad:2015ova,Lesgourgues:2015wza,Buen-Abad:2017gxg,Krall:2017xcw,Pan:2018zha}). Since these models tend to predict strong self-interactions in the DR sector, we will always assume in this section that DR is a relativistic perfect fluid described by one continuity and one Euler equation (unlike for $n=4$ and $n=2$). To stress this difference, we will denote the DR density (in units of effective neutrino number) $\Delta N_\mathrm{fluid}$ instead of $\Delta N_\mathrm{eff}$.

For the $n=0$ model, we can use different parametrisations and priors corresponding to different approaches discussed in the previous literature -- either in the ETHOS general framework, or for specific models like the NADM one. We first look at the standard ETHOS parametrisation, with the same choice of priors as in previous cases. Then, to compare our results with Refs.~\cite{Lesgourgues:2015wza,Buen-Abad:2017gxg}, we will switch to linear priors on the interaction rate combined with either linear or logarithmic priors on the parameters  $\Delta N_\mathrm{fluid}$. This will also allow us to see the influence of the choice of priors on our results.

\vspace{0.5cm}

\noindent {\it CMB constraints with ETHOS $n=0$ parametrisation.} Our results for this case, assuming a flat prior on $\xi\geq 0$ and on $-3 \leq \log_{10}(a_\mathrm{dark}/\mathrm{Mpc}^{-1}) \leq 20$, are shown in Fig.~\ref{fig:n0} and Table~\ref{table:n0}.
In this case, the general behaviour is similar to the previous cases: we obtain an upper bound of $\xi$ and a hyperbolic-shaped upper limit on ($\xi, a_\mathrm{dark}$). CMB bounds are much stronger in this model than in previous cases, which is consistent with the fact that the rate $\Gamma_\mathrm{DM-DR}(z)$ evaluated near photon decoupling, when $z \sim {\cal O}(10^3)$, is much larger for the same value of $a_\mathrm{dark} \xi^4$ when $n$ decreases (as shown by equation~(\ref{eq:gammadmdrz})). We shall see that for $n=0$, CMB bounds dominate over Lyman-$\alpha$ bounds at least for some values of $\xi$. Thus it is worth quantifying these bounds precisely. In the space ($\xi, \log_{10}(a_\mathrm{dark})$) and within our prior range, the 95~\%~C.L. preferred region is defined in good approximation by:
\begin{itemize}
\item either $\xi<0.13$,
\item or $\xi < 0.38$ and $10^4 a_\mathrm{dark} \xi^4 < 14 \, \mathrm{Mpc}^{-1}$.
\end{itemize}
This means that the CMB excludes all ETHOS $n=0$ models with either a too large DR density  ($\xi>0.38$) or a too large $\Gamma_\mathrm{DM-DR}$ rate ($10^4 a_\mathrm{dark} \xi^4 > 14  \, \mathrm{Mpc}^{-1}$), but looses sensitivity to these parameters when the DR density is very small ($\xi<0.13$). 

\begin{table}[t]
\begin{center}
\begingroup
\renewcommand{\arraystretch}{1.5} 
\begin{tabular}{|c|c|c|c|}
\hline
& $\Lambda$CDM & \multicolumn{2}{c|}{ETHOS $n=0$}\\ 
parameter & Planck + BAO & Planck + BAO & + Lyman-$\alpha$\\
\hline \hline
$100~\omega_\rr{b }$ & $2.219_{-0.014}^{+0.013}$ & $2.220_{-0.015}^{+0.015}$   & $2.221_{-0.015}^{+0.015}$ \\ 
$\omega_\rr{cdm }$ & $0.1192_{-0.0010}^{+0.0011}$ & $0.1195_{-0.0014}^{+0.0011}$ & $0.1192_{-0.001}^{+0.001}$ \\ 
$\log(10^{10}A_\rr{s })$ & $3.050_{-0.023}^{+0.023}$ & $3.053_{-0.024}^{+0.025}$  & $3.054_{-0.024}^{+0.025}$\\ 
$n_\rr{s }$ & $0.9618_{-0.0041}^{+0.0042}$ & $0.9621_{-0.0045}^{+0.0042}$ & $0.9624_{-0.0041}^{+0.0044}$\\ 
$\tau_\rr{reio }$ & $0.060_{-0.012}^{+0.012}$ & $0.061_{-0.012}^{+0.013}$  & $0.061_{-0.014}^{+0.013}$ \\ 
$H_0 \ / \left[{\rm km/(s \,Mpc)}\right]$ & $67.94_{-0.49}^{+0.46}$ & $68.04_{-0.60}^{+0.50}$  &$68.03_{-0.49}^{+0.47}$\\ 
$\sigma_8$ & $0.8234_{-0.0090}^{+0.0085}$ & $0.815_{-0.009}^{+0.044}$ & $0.8237_{-0.0093}^{+0.0097}$\\
$n_\rr{eff}$ & $-2.308_{-0.0035}^{+0.0034}$ & $-3.4_{-4.2}^{+9.5}$  & $-2.3100_{-0.0079}^{+0.0071}$ \\
$\xi$ & -- & 
$<0.38$ & 
$<0.33$ \\ 
$\log_{10}(a_\rr{dark } \ / \left[{\rm Mpc}^{-1}\right])$ & -- & n.l. & $ < 3.3$\\ 
$\Delta\chi^2$ & -- & $0$ & $-0.70$\\
\hline
\hline
$\Delta N_\mathrm{fluid}$ & -- & -- &$<0.47$ \\
$10^{4} a_\mathrm{dark}\xi^4/   \left[{\rm Mpc}^{-1}\right]$ & -- & -- &$<14$ \\
\hline
\end{tabular}
\endgroup
\end{center}
\caption{Parameter 68\,\% confidence limits (or 95\,\% upper bound in some cases) for all relevant parameters for the ETHOS $ n = 0 $ case, both with Planck + BAO and Planck + BAO + Lyman-$\alpha$. 
With the first dataset, the interaction parameter is not bounded within the prior range. The $\Delta\chi^2$ is given with respect to $\Lambda$CDM with the same datasets. The last two rows show the results obtained with linear priors on $\Delta N_\mathrm{fluid}$ and $10^4a_\mathrm{dark}\xi^4$ using the second dataset.}
\label{table:n0}
\end{table}

Another interesting aspect of these results is that the ETHOS $n=0$ model allows to reach larger values of $H_0$ or lower values of $\sigma_8$ than the $\Lambda$CDM model. By looking at the two-dimensional contour plots in the left panel of Fig.~\ref{fig:n0}, we see that:
\begin{itemize}
\item high values of $H_0$ require  a large DR density, $\xi>0.4$ (i.e. $\Delta N_\mathrm{fluid}>0.1$): indeed this is a consequence of the well-known $H_0 -  \Delta N_\mathrm{fluid}$ degeneracy, that works particularly well in this case because DR is self-interacting, and thus less constrained by CMB observables than extra free-streaming relics \cite{Lesgourgues:2015wza}. Our 95~\%~CL preferred region reaches values up to $H_0\simeq 70\,\mathrm{km} \,\mathrm{s}^{-1} \mathrm{Mpc}^{-1}$ for $\xi \simeq 0.38$ ( $\Delta N_\mathrm{fluid}\simeq 0.08$). With our choice of priors, this part of the allowed parameter space has little weight, and the 68~\%~C.L. preferred interval for $H_0$ is still nearly the same as for $\Lambda$CDM. Later in this section, runs with different priors will give more weight to this degeneracy.
\item a close inspection of the  ($\xi, \sigma_8$) contour plot of Fig.~\ref{fig:n0} shows that in this plane, the marginalised posterior is bimodal, i.e. made of the superposition of two separate categories of models. The first one has  $\sigma_8 = 0.823 \pm 0.017$ (95~\%~C.L.) for any allowed value of the DR density parameter (in the range $0<\xi<0.38$). The second one corresponds to a strongly degenerate direction in  ($\xi, \sigma_8$), captured by the relation $\sigma_8 \simeq 0.823-210\, \xi^4$ (like for the ETHOS $n=2$ model), and requires a large interaction rate $a_\mathrm{dark} \geq 1$. It stretches down to $\sigma_8 = 0.68$ for $\xi \simeq 0.16$.
This part of the parameter space will also play an enhanced role in some of the runs that we will perform later with different physical motivations and priors. 
\end{itemize}

\vspace{0.5cm}

\noindent {\it Lyman-$\alpha$ constraints with ETHOS $n=0$ parametrisation.} At first sight, the discussion of the Lyman-$\alpha$ constraints seems very similar to that for $n=2$ or 4. We expect that Lyman-$\alpha$ data will slightly tighten the bound on $\xi$ and put a strong limit on $10^4 a_\mathrm{dark} \xi^4 < {\cal O}(10)$. This is indeed what happens in our run with a linear prior on $\xi$ and a logarithmic prior on $a_\mathrm{dark}$: we get $\xi<0.33$ and $10^4 a_\mathrm{dark}\xi^4 < 14 \, \mathrm{Mpc}^{-1}$ (95~\%~C.L.). Doing a second run with flat priors on ($\Delta N_\mathrm{fluid}$, $10^4 a_\mathrm{dark}\xi^4$), we find $\Delta N_\mathrm{fluid}< 0.47$ and a confirmation of $10^4 a_\mathrm{dark}\xi^4 < 14 \, \mathrm{Mpc}^{-1}$ (95~\%~C.L.).

However, a run with the ``Planck + BAO + Lyman-$\alpha$ Prior'' combination shows that the previous results must be taken with great care. Looking at the middle plot of the left panel of Fig.~\ref{fig:n0}, we see that:
\begin{itemize}
\item the different checks performed inside our Lyman-$\alpha$ likelihood induce a cut at $\xi<0.33$. Thus the previous bound on $\xi$ did not come from the Lyman-$\alpha$ data but from our methodology, i.e. from the fact that ETHOS $n=0$ models with $\xi>0.33$ do not yield a power spectrum that can be accurately represented by the $\{\alpha, \beta, \gamma \}$-parametrisation.  Thus we should not trust any bound on $\xi$ apart from the one obtained with Planck+BAO alone, namely $\xi<0.38$. 
\item for $\xi>0.13$, the upper bound on $a_\mathrm{dark}\xi^4$ is nearly the same in the three ETHOS $n=0$  runs (without Lyman-$\alpha$ likelihood, with Lyman-$\alpha$ Prior and with Lyman-$\alpha$ Data), suggesting that CMB data alone provide the strongest bounds in this case: $10^4 a_\mathrm{dark}\xi^4 < 14\, \mathrm{Mpc}^{-1}$ (95~\%~C.L.). Given the impact of this model on CMB and LSS observables, already discussed in previous works~\cite{Lesgourgues:2015wza,Buen-Abad:2017gxg}, this is not a surprise: for parameter values leading to significant effects in the CMB temperature and polarisation spectrum, this model only generates a very smooth and progressive suppression in the small-scale matter power spectrum, much more difficult to constrain with Lyman-$\alpha$ data that the sharp exponential cut-off observed for $n=2,4$.
\item for $\xi<0.02$, the Lyman-$\alpha$ Prior run sets no upper limit on the interaction rate, while the Lyman-$\alpha$ Data run returns $10^4 a_\mathrm{dark}\xi^4 < 14$ (95~\%~C.L.): thus we can trust this bound which really comes from the data.
\item there is a problematic range $0.02<\xi<0.13$ in which the Lyman-$\alpha$ Prior run also sets an upper limit $10^4 a_\mathrm{dark}\xi^4 < 14\, \mathrm{Mpc}^{-1}$. The reason is that for $n=0$ and $\xi>0.02$, the $\{\alpha, \beta, \gamma \}$ parametric function cannot provide an accurate fit of the suppression in the matter power spectrum. This indicates that for this class of models, the bounds are driven by the limitations of the method, in particular by the flexibility of the parametric fitting function, and not by the data.
We could search for a better method, but we believe that this is not well motivated, for two reasons.
First, $0.02<\xi<0.13$ means $6\cdot 10^{-7} < \Delta N_\mathrm{fluid} < 10^{-3}$. The weight of this region would be negligible if we would run with a flat prior on $\Delta N_\mathrm{fluid}$, so we may simply ignore it. Second, the analytic argument suggesting that the Lyman-$\alpha$ bound on the DM-DR interaction takes the form of an upper limit on $a_\mathrm{dark}\xi^4$ worked very well for $n=2$ and $n=4$, and still works very well in the present case for $\xi<0.02$ and $\xi>0.13$. We have no reason to believe that this would not be the case in the intermediate range. Thus it is reasonable to expect that a better method would return $10^4 a_\mathrm{dark}\xi^4 < 14$ (95~\%~C.L.) throughout the range of allowed values $0<\xi<0.38$.
\end{itemize}
In summary, we should retain from this analysis that, for $\xi>0.13$ Lyman-$\alpha$ data, at least with our approach, cannot improve over Planck + BAO bounds, which give $10^4 a_\mathrm{dark}\xi^4 < 14\, \mathrm{Mpc}^{-1}$ (95~\%~C.L.). For $0<\xi<0.02$, the Lyman-$\alpha$ data give the same bound. In the intermediate range, a different approach would be needed, but there are some hints that the Lyman-$\alpha$ data would give again the same bound.

\noindent {\it CMB constraints with a particle-physics-motivated flat prior on $\Delta N_\mathrm{fluid}\geq 0.07$.} Several works have presented particle physics models that can be effectively described by the ETHOS $n=0$ parametrisation, with weakly interacting DM-DR, and strongly self-interacting DR. In the NADM model~\cite{Buen-Abad:2015ova}, the DR is made up of the dark gluons of a non-abelian gauge symmetry $SU(N)$. Its density is parametrised by $\Delta N_\mathrm{fluid} = 0.07 (N^2-1)$ with $N \geq 2$. Ref.~\cite{Lesgourgues:2015wza} presents a second set-up leading to approximately the same cosmological signature, in which the DR has two components: the dark photon of a dark $U(1)$ gauge symmetry, plus $N_f$ massless fermions with a dark charge $q$. For $q\geq1/3$ the DR density is parametrised by  $\Delta N_\mathrm{fluid} = 0.07 (1+\frac{7}{4} N_f)$, but for smaller charges one gets $\Delta N_\mathrm{fluid} = 0.07$. These models motivate dedicated runs with a flat prior on $\Delta N_\mathrm{fluid}\geq 0.07$. To compare our results with previous works, we will also adopt a flat prior on the DM-DR momentum exchange rate evaluated today, $\Gamma_0$, related to the ETHOS parameters through 
$a_0 \Gamma_0 = \Gamma_\mathrm{DM-DR}(z=0)=\frac{4}{3} \omega_\mathrm{DR} \, a_\mathrm{dark}=\frac{4}{3} \omega_\gamma \, a_\mathrm{dark}\xi^4$. 
With such a correspondence, we checked that we could accurately reproduce Figs. 3-6 of Ref.~\cite{Buen-Abad:2017gxg}: thus  our version of {\sc class} modified for the ETHOS parametrisation does agree perfectly with the version of {\sc class} modified specifically for the NADM model in Ref.~\cite{Buen-Abad:2017gxg}.

%

The prior $\Delta N_\mathrm{fluid} \geq 0.07$ translates in the ETHOS parametrisation to $\xi\geq 0.367$. Looking at our previous results, we see that this clearly corresponds to the region in which the CMB bounds are at least as strong as the Lyman-$\alpha$ bounds: thus for this case it is sufficient to run with Planck + BAO data only. Note that with such a prior, we avoid the bi-modality of the posterior found in the results of the previous run (corresponding to a degeneracy between $\sigma_8$ and $\xi$ for $\xi\leq0.16$). Thus the theoretical prior $\Delta N_\mathrm{fluid} \geq 0.07$ offers a technical advantage: it limits the exploration of the model parameter space to a region where the posterior is smooth and unimodal, leading to more robust MCMC results.

\begin{figure}[t!]
\begin{center}
\begin{tabular}{cc}
\includegraphics[width=0.45\linewidth]{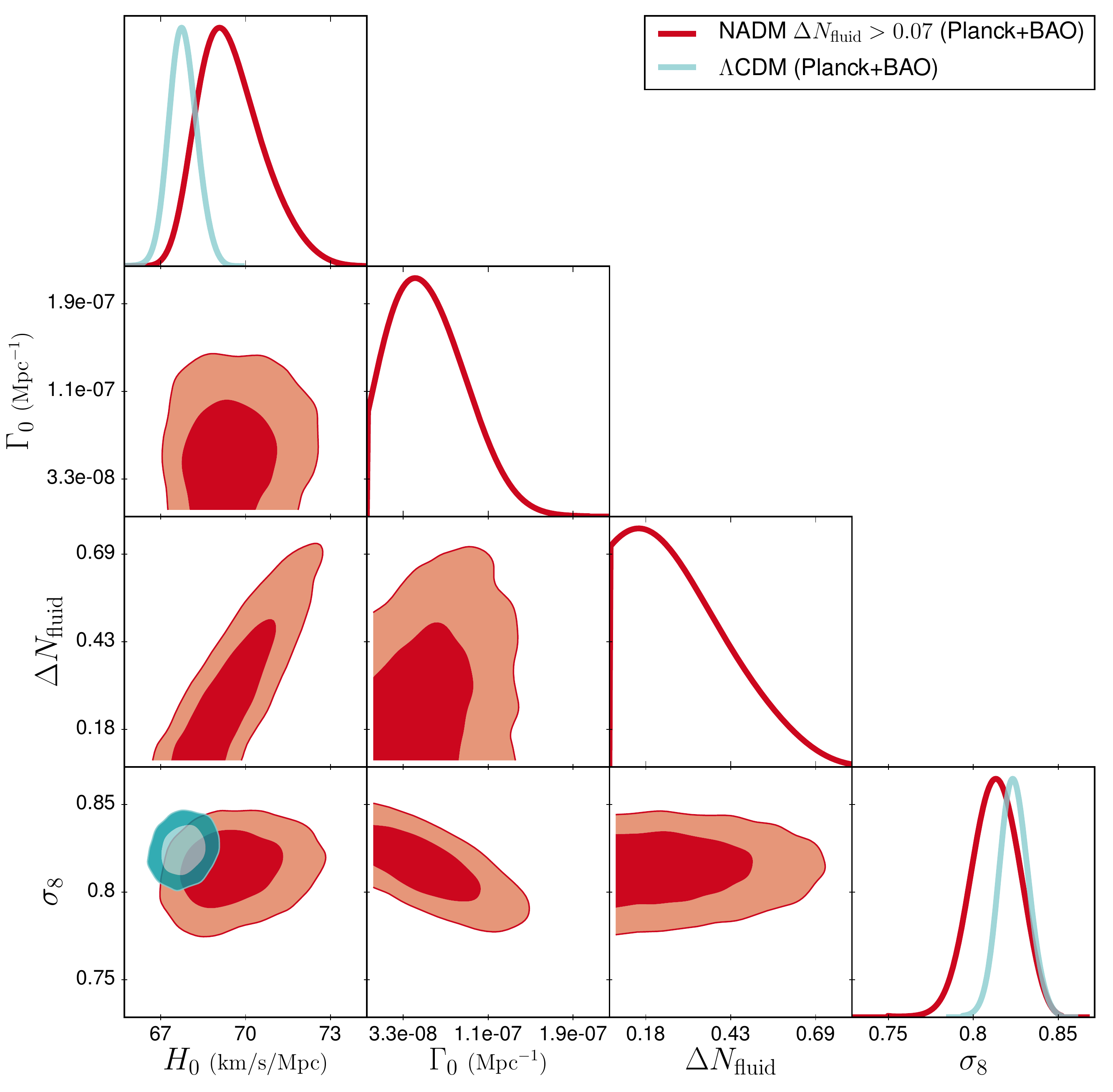}&
\includegraphics[width=0.45\linewidth]{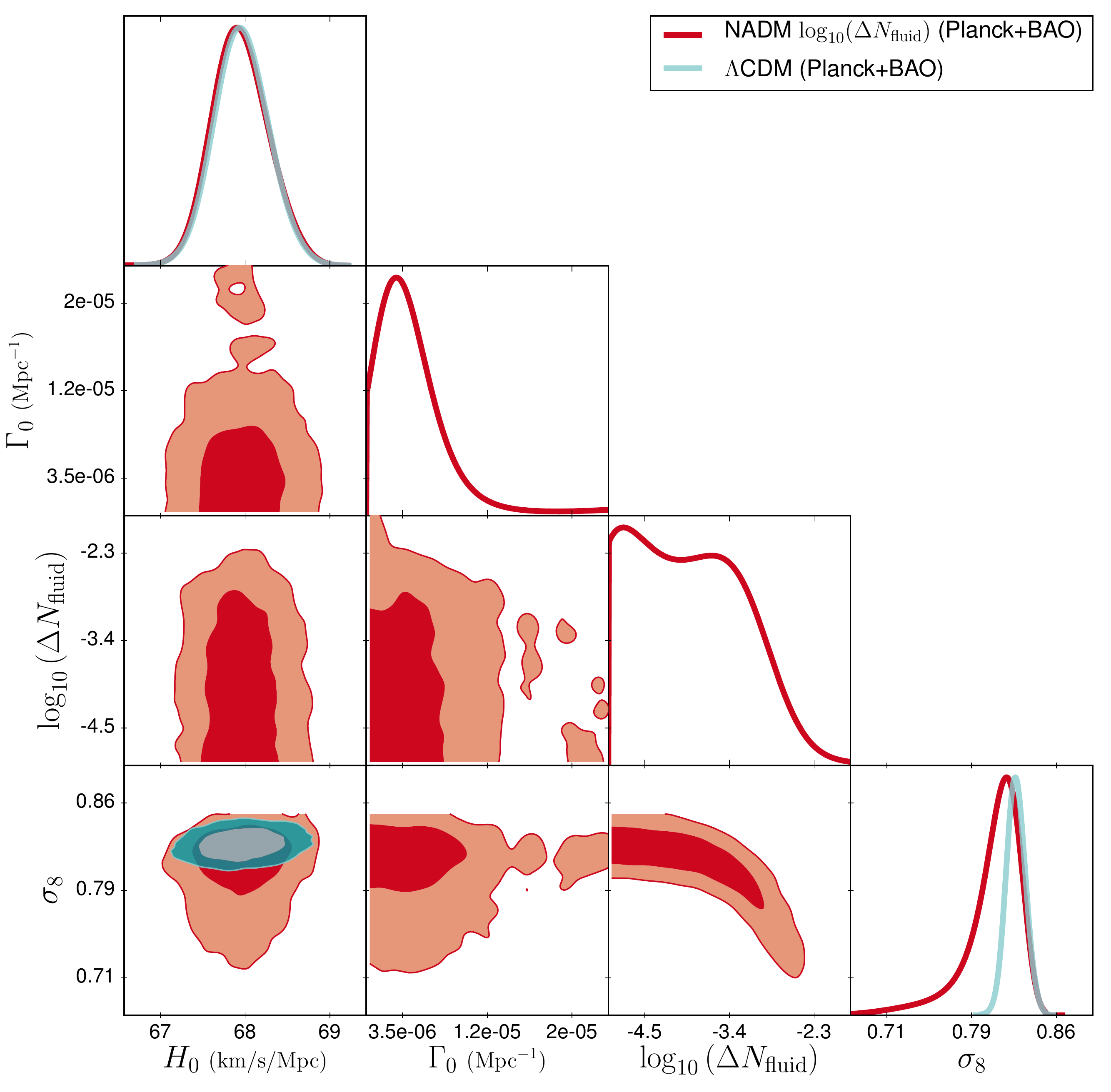}
\end{tabular}
\caption{\emph{(Left)} Two-dimensional posterior distributions for all main parameters using Planck + BAO, for the NADM case (red) and for $\Lambda$CDM (blue),  with the lower prior $\Delta N_\mathrm{fluid}>0.07$. \emph{(Right)} Same as left but with the log prior $-5 \leq \log_{10}(\Delta N_\mathrm{fluid}) \leq 0$. }
\label{fig:nadm}
\end{center}
\end{figure}

Our findings, presented in the left panel of Fig.~\ref{fig:nadm} and middle column of Table~\ref{table:nadm}, are consistent with those of Refs.~\cite{Lesgourgues:2015wza,Krall:2017xcw} when using Planck 2015 + BAO 2011 data. Our bounds are however slightly stronger and more up-to-date, because we include  Planck lensing data and more recent BAO data. We do not compare directly our results with those of Ref.~\cite{Buen-Abad:2017gxg}, as the latter always included direct $H_0$ measurements, as well as  Planck data on Sunyaev-Zel'dovitch cluster counts.

We find  $0.07 \leq \Delta N_\mathrm{fluid} \leq 0.59$ (95~\%~C.L.), corresponding to $0.367 < \xi < 0.626$ with a non-flat prior on $\xi$, and $\Gamma_0 < 1.2 \cdot 10^{-7} \mathrm{Mpc}^{-1}$ (95~\%~C.L.), corresponding to $10^4 a_\mathrm{dark} \xi^4 < 36$ (95~\%~C.L.). We see that the lower prior edge on $\Delta N_\mathrm{fluid}$ and the linear prior on both $\Delta N_\mathrm{fluid}$  and $\Gamma_0$ have pushed the MCMC to explore regions that were not reached with the previous ETHOS $n=0$ prior: the previous preferred region only stretched up to twice smaller values of $\xi$ and $10^4 a_\mathrm{dark} \xi^4$. However, the current run is not forced to explore a region in tension with the data, since the best-fit $\chi^2$ only increases marginally (by 1.9) with respect to the best-fit $\Lambda$CDM $\chi^2$.

Even if this model is not preferred by Planck + BAO data, it remains very interesting as a possible way to reconcile CMB+BAO data with high values of $H_0$ and low values of $\sigma_8$~\cite{Lesgourgues:2015wza,Buen-Abad:2017gxg}. Indeed, we find that this model can accommodate a large  $H_0 = 69.6_{-1.3}^{+0.8}$ (68~\%~C.L.) reducing the tension with the most recent SH0ES data \cite{Riess2019} from 4.1$\sigma$ to 2.7$\sigma$, and a low $\sigma_8=0.813_{-0.012}^{+0.015}$ (68~\%~C.L.). It also allows for smaller values of the parameter combination $S_8 \equiv \sigma_8 \sqrt{\Omega_\mathrm{m}/0.3} = 0.813_{-0.012}^{+0.015}$ (68~\%~C.L.)\footnote{For this model, we find exactly the same bounds on $\sigma_8$ and $S_8$, because $\Omega_\mathrm{m}$ remains very close to 0.3.} than the $\Lambda$CDM model which gives $S_8 = 0.8235_{-0.0091}^{+0.0088}$ (68~\%~C.L.) for the same dataset. Thus it increases the compatibility with the combined KiDS + VIKING-450 + DES-Y1 measurement of Ref.~\cite{Joudaki:2019pmv} from $2.3\sigma$ to $1.8\sigma$ level\footnote{Measurements of $S_8$ from weak lensing surveys are still very debated and potentially affected by poorly known systematics; for instance, the independent analysis of Ref.~\cite{Fluri:2019qtp} gives a result compatible with our $\Lambda$CDM $S_8$ bounds at the 1.2$\sigma$ level.}. The physical explanation is that this model is able to exploit the $H_0 - \Delta N_\mathrm{fluid}$ degeneracy thanks to its self-interacting DR component, while reducing at the same time the small-scale matter power spectrum amplitude  thanks to the effect of DR dragging DM perturbations.

%
%
%

\begin{table}[t!]
\begin{center}
\begingroup
\renewcommand{\arraystretch}{1.5} 
\begin{tabular}{|c||c|c|c|}
\hline
parameter &  $\Lambda$CDM & $\Gamma_0>0$, $\Delta N_\rr{fluid}>0.07$  &  $\Gamma_0>0$, $-5 \leq \log(\Delta N_\rr{fluid}) \leq 0$ \\ 
\hline \hline
$100~\omega_\rr{b }$ & $2.219_{-0.014}^{+0.013}$ & $2.232_{-0.019}^{+0.017}$&$2.219_{-0.016}^{+0.014}$ \\
$\omega_\rr{cdm }$ & $0.1192_{-0.0010}^{+0.0011}$ & $0.1249_{-0.0037}^{+0.0023}$&$0.1192_{-0.0011}^{+0.0011}$ \\
$ln10^{10}A_\rr{s }$ & $3.050_{-0.023}^{+0.023}$ & $3.069_{-0.025}^{+0.026}$ & $3.054_{-0.026}^{+0.025}$\\
$n_{s }$ & $0.9618_{-0.0041}^{+0.0042}$ & $0.9653_{-0.0045}^{+0.0042}$&  $0.9617_{-0.0045}^{+0.0042}$\\
$\tau_\rr{reio }$ & $0.060_{-0.012}^{+0.012}$ & $0.0696_{-0.013}^{+0.013}$ & $0.06181_{-0.014}^{+0.013}$ \\
$H_0 \ / \left[{\rm km/(s \,Mpc)}\right]$ & $67.94_{-0.49}^{+0.46}$ & $69.55_{-1.3}^{+0.84}$&  $67.94_{-0.50}^{+0.48}$ \\
$\sigma_8$ & $0.8234_{-0.0090}^{+0.0085}$ & $0.813_{-0.012}^{+0.015}$ &$0.806_{-0.011}^{+0.029}$ \\
$n_\rr{eff}$ & $-2.308_{-0.0035}^{+0.0034}$ & $-2.332_{-0.011}^{+0.018}$ &$-3.261_{-0.36}^{+0.96}$\\
$\Gamma_{0} \ / \left[{\rm Mpc}^{-1}\right]$ & -- & $<1.2\cdot 10^{-7}$ &$<1.5\cdot 10^{-5}$ \\
$\Delta N_\mathrm{fluid}$ & -- & $<0.59$ & --  \\
$\log_{10}(\Delta N_\mathrm{fluid})$ & -- & -- &$<-2.66$ \\
$\Delta\chi^2$ & -- & $1.90$ & $2.34$ \\
\hline
\end{tabular}
\endgroup
\end{center}
\caption{Parameter 68\,\% confidence limits (or 95\,\% upper bound in some cases) for all relevant parameters for the NADM case with two different prior choices, and using Planck + BAO. The $\Delta\chi^2$ is given with respect to $\Lambda$CDM with the same datasets.}
\label{table:nadm}
\end{table}

These results are consistent with those based on the previous ETHOS $n=0$ parametrisation (with flat priors on $\xi$ and $\log_{10}(a_\mathrm{dark})$ and the same dataset), although the comparison is not straightforward since the new run explores a different region of the parameter space. The previous results did show the trend to accommodate a larger $H_0$ when $\Delta N_\mathrm{fluid}$ increases. This is even clearer in this run that reaches higher values of $\Delta N_\mathrm{fluid}$. The previous results also showed that when the interaction rate increases from $\log_{10}(a_\mathrm{dark}) \simeq -2$ to $\log_{10}(a_\mathrm{dark}) \simeq 0$, smaller values of $\sigma_8$ can be reached. This is confirmed in the new run by the clear correlation between the interaction rate and $\sigma_8$ in the left panel of Fig.~\ref{fig:nadm}. 

The comparison between the two runs alerts us on the fact that the ability of this model to reconcile datasets depends on the priors: the model would appear less effective in this respect with a lower prior edge $\Delta N_\mathrm{fluid} \geq 0$ (or with logarithmic priors on  $\Delta N_\mathrm{fluid}$ or $\Gamma_0$). This prior dependence of the conclusions applies anyway to most of the models attempting to resolve the tensions, and would only go away if we included the anomalous $H_0$ and $\sigma_8$ data in the analysis: then, even with different priors, some non-zero values of $\Delta N_\mathrm{fluid}$ and $\Gamma_0$ would be preferred with a statistical significance of a few sigmas.

\noindent {\it CMB constraints with a logarithmic prior on $\Delta N_\mathrm{fluid}$.} Ref.~\cite{Buen-Abad:2017gxg} explored the same model with a flat prior on $-5 \leq \log_{10}(\Delta N_\rr{fluid}) \leq 0$ and on $\Gamma_0\geq0$. The motivation for this prior was to provide complementary results to the previous case, exploring very small values of the DR density which can always be motivated by specific particle physics constructions. We will now update these results with our Planck + BAO dataset, still not using Lyman-$\alpha$ data here, as we have seen that our method cannot provide accurate constraints for these models. 

Our results are presented in the right panel of Fig.~\ref{fig:nadm} and right column of Table~\ref{table:nadm}. We find $-5 \leq \log_{10} \Delta N_\mathrm{fluid} \leq -2.66$ (95~\%~C.L.), corresponding to $0.04 < \xi < 0.15$ with a non-flat prior on $\xi$, and $\Gamma_0 < 1.5 \cdot 10^{-5} \mathrm{Mpc}^{-1}$ (95~\%~C.L.), corresponding to $a_\mathrm{dark} \xi^4 < 0.45$ (95~\%~C.L.). With this prior choice, we no longer allow for larger $H_0$, which is in agreement with our ETHOS $n=0$ results. This can be understood in the following way: the flat prior on $\xi$ (and indeed the log prior on $\Delta N_\rr{fluid}$) gives less weight to large amounts of DR, and thus the possibility to relax the $H_0$ tension goes away. However,  we can accommodate lower $\sigma_8$, thanks to a degeneracy between $\sigma_8$ and $\log_{10} \Delta N_\mathrm{fluid}$ that is clearly visible in the right panel of Fig.~\ref{fig:nadm}. This degeneracy is equivalent to the $\sigma_8-\xi$ degeneracy previously observed in the ETHOS $n=0$ results, and could in principle reconcile the Planck + BAO data with values as low as $\sigma_8 \sim 0.7$.  The model predicts $S_8= 0.8058_{-0.0085}^{+0.0088}$ (68~\%~C.L.), which is compatible with KiDS + VIKING-450 + DES-Y1~\cite{Joudaki:2019pmv} at the $1.7\sigma$ level.

%% file: Conclusions.tex
\section{Discussion and conclusions}
\label{sec:conc}

The small scale crisis of collisionless CDM and the lack of a detection of WIMPs open up to theoretical models featuring a non-collisionless behaviour of Dark Matter particles that interact with a non-standard relativistic component, Dark Radiation. We implemented these models in {\sc class}~\footnote{The modified version of {\sc class} will be made available upon acceptance of the paper.} following the effective ETHOS parametrisation, which is flexible enough to reproduce the impact on cosmological observables at linear scales of any particle physics model leading to a drag force between DM and DR.

Typically models of DM-DR interactions show peculiar features in the matter power spectrum at very small scales. Thus, Lyman-$\alpha$ data represent an excellent probe to constrain them. However, linear perturbation theory fails to reproduce clustering on such small scales. Applying the usual non-linear corrections derived in a CDM framework might lead to incorrect results. In this paper we overcome this issue by constructing a likelihood~\footnote{Our Lyman-$\alpha$ likelihood will be released once the paper will be accepted.} in {\sc MontePython} based on the method proposed in Ref.~\cite{Murgia:2017lwo, Murgia:2017cvj, Murgia:2018now}, that maps the suppression of the linear matter power spectrum into a parametric function depending only on three effective parameters $\{\alpha, \beta, \gamma\}$. For each tuple the likelihood interpolates among the Lyman-$\alpha$ $\chi^2$ computed on the nodes of a grid in $\{\alpha, \beta, \gamma\}$ for which hydrodynamical simulations were run, to produce the flux power spectrum  to compare to the MIKE/HIRES data.
Therefore, the constraints on DM-DR interactions derived with our pipeline are related to the true observable probed by up-to-date Lyman-$\alpha$ data, i.e. the flux power spectrum, rather than to the (model dependent) inferred amplitude, slope and curvature of the linear matter power spectrum at the scales probed by the forest~\cite{Krall:2017xcw}. 

We applied our method to three different cases, corresponding to three different scaling relations of the comoving DM-DR interaction rate with respect to the temperature. For each case we performed MCMC runs for two dataset combinations (Planck+BAO and adding Lyman-$\alpha$) and we studied the impact of different prior assumptions on the relevant DM-DR parameters.

\vspace{0.3cm}
\noindent{\it Temperature dependent comoving interaction rate: $n=4$ and $n=2$.}
The Planck+BAO constraints in the DM-DR parameter space ($\log_{10}(a_\rr{dark}/\rr{Mpc}^{-1})$ versus $\xi$) show an hyperbolic behaviour indicating a degeneracy between small (large) amount of DR and large (small) interaction strength. The $2\sigma$ upper bounds on the amount of DR ($n=4:\,\xi<0.40$ and $n=2:\,\xi<0.43$ at 95\% C.L.) come from the impact on the CMB power spectra of a combination of effects: the presence of DR itself and the induced drag force on DM. On the other hand, Planck+BAO do not provide a unique upper bound on $\log_{10}(a_\rr{dark}/\rr{Mpc}^{-1})$ because of the aforementioned degeneracy leading to an asymptote in $\log_{10}(a_\rr{dark}/\rr{Mpc}^{-1})$ at small $\xi$.

\noindent When applying the Lyman-$\alpha$ likelihood, first of all we checked that the $\{\alpha, \beta, \gamma\}$ parametric function succesfully reproduce the suppression of the matter power spectrum for most of the parameter space $(\log_{10}(a_\rr{dark}/\rr{Mpc}^{-1}), \xi)$. The only exception is a tiny region at $\xi<0.05$ (and only when $n=2$). This corresponds to $\Delta N_\mathrm{eff} < 2 \cdot 10^{-5}$, a range difficult to motivate theoretically since it would normally correspond to a number of degrees of freedom at DR decoupling a few orders of magnitude larger than the SM and MSSM expectations. 
While the inclusion of the Lyman-$\alpha$ data only slightly improves the bounds on DR ($n=4:\,\xi<0.38$ and $n=2:\,\xi<0.40$ at 95\% C.L.), it is essential to set upper limits on the interaction strength.
We found that the data fixes a limit on the parameter combination $a_\rr{dark} \xi^4$ that fixes the normalization of the DM-DR interaction rate $\Gamma_\mathrm{DM-DR}$.


\noindent In order to further investigate this behaviour, we performed new runs with flat priors on different parameters: $\Delta N_\rr{eff}$ and the combination $10^{n-4}(a_\rr{dark}/\rr{Mpc}^{-1})\xi^4$. This choice is motivated by the fact that the front factor of the DM dipole moment is proportional to $(a_\rr{dark}/\rr{Mpc}^{-1})\xi^4$, and, thus, it is the relevant quantity for determining the drag epoch. Given the flat prior on $\Delta N_\rr{eff}$, the bounds on the amount of DR are less constraining than those derived with a flat prior on $\xi$: $\Delta N_\rr{eff}<0.23$ for $n=4$ and $\Delta N_\rr{eff}<0.29$ for $n=2$ at 95\% C.L..
The limits obtained on the parameter combination $10^{4-n}(a_\rr{dark}/\rr{Mpc}^{-1})\xi^4$ ($<30$ for $n=4$ and $<18$ for $n=2$ at 95\% C.L.) can then be translated into constraints on the actual particle physics model and, thus, on the impact on the small scale crisis. For instance, in the $n=4$ case, the bound still leaves room for a cut-off mass in the halo mass function that can solve the missing satellite problem.

\noindent Concerning the infamous cosmological tensions, the constraints on $H_0$ and on $\sigma_8$ are consistent with $\Lambda$CDM, thus, once Lyman-$\alpha$ data are included, none of these models alleviate the discrepancies. Finally, the $\chi^2$ analysis shows that the global fit provided by DM-DR models is comparable with the one of $\Lambda$CDM.

\vspace{0.3cm}
\noindent{\it Constant comoving interaction rate: $n=0$.}
Using the ETHOS parametrisation, the results look similar to those of the previous cases. The shape of the limits in the $(\log_{10}(a_\rr{dark}/\rr{Mpc}^{-1}), \xi)$ plane is hyperbolic. Planck+BAO constrain only $\xi<0.38$ at 95\% C.L.. The inclusion of Lyman-$\alpha$ slightly improves the constraints on $\xi$ ($\xi<0.33$) and sets an upper bound $\log_{10}(a_\rr{dark}/\rr{Mpc}^{-1})<3.3$. And with different priors, we obtained $\Delta N_\rr{eff}<0.47$ and $10^{4}a_\rr{dark}\xi^4<14\,\rr{Mpc}^{-1}$ at 95\% C.L..
However, an accurate analysis showed that this case needs a dedicated discussion.
Indeed, above a certain threshold in $\xi$ ($\xi \gtrsim 0.13$) the dominant constraints come from Planck. This was expected since, for a given value of $a_\rr{dark}$, the impact on the CMB is more pronounced in the $n=0$ case than in the case of a temperature dependent comoving interaction rate. We also noticed that for large values of $\xi$ the Lyman-$\alpha$ likelihood cannot be applied, because the $\{\alpha, \beta, \gamma\}$ parametric function cannot reproduce the smooth suppression of the matter power spectrum realised by these models.

\noindent We further investigated this case by fitting only Planck+BAO and using the same parametrisation as Ref.~\cite{Buen-Abad:2017gxg}, i.e. $\Delta N_\rr{fluid}$ and $\Gamma_0$. The various dark sector set-ups discussed in Refs.~\cite{Buen-Abad:2015ova,Lesgourgues:2015wza} motivate a flat prior on $\Delta N_\rr{fluid}>0.07$, which leads to the 95\% C.L. upper bounds $\Delta N_\rr{fluid}<0.59$ and $\Gamma_0<1.2 \cdot 10^{-7}\,\rr{Mpc}^{-1}$. The prior opens up to a larger amount of DR, as well as a stronger interaction rate, thus, inducing a reduction of the $H_0$ tension from $4.1\sigma$ to $2.7\sigma$, and a mitigation of the $\sigma_8$ tension from $2.3\sigma$ to $1.8\sigma$. However, once a flat prior on $-5 \leq \log_{10}(\Delta N_\rr{fluid})\leq 0$ is assumed, the former tension is restored, while the latter is still mitigated ($1.7\sigma$).

\vspace{0.3cm}
\noindent{\it Conclusions.}
Our analysis showed that there is still room for solving the cosmological and astrophysical tensions by means of a modified dark sector devising interactions between the non-relativistic DM and a new relativistic component.
However, before claiming a solution of the $H_0$ and $\sigma_8$ tensions, a careful analysis of the prior dependence of the results must be performed, paying specific attention to the quantities relevant for constraining the interactions, i.e. the drag force. On the other hand, concerning the astrophysical tensions,
our Lyman-$\alpha$ likelihood implemented in {\sc MontePython} provides an efficient tool to investigate models featuring a suppression in the matter power spectrum at non-linear scales, and can be applied to scenarios beyond the cases investigated in the present paper. The bounds derived with this approach are robust and can boost the speculation about DM models.